\documentclass{aastex}
\usepackage{enumerate}
\usepackage{float}
\usepackage{placeins}

\newcommand{\degree}{\ensuremath{^\circ}}
\newcommand{\emaila}{david@mso.anu.edu.au}
\newcommand{\simlt}%
        {\,\hbox{\lower0.5ex\hbox{$\sim$}\llap{\raise0.5ex\hbox{$<$}}}\,}
\newcommand{\Lsolar}{\mbox{\,$\rm L_{\odot}$}}        
\newcommand{\Msolar}{\mbox{\,$\rm M_{\odot}$}}        

\begin{document}

\title{The Small Isolated Gas Rich Irregular Dwarf (SIGRID) Galaxy Sample: Description and First Results}
\slugcomment{Submitted to Astronomical Journal}

\shorttitle{SIGRID galaxy sample - Description}
\shortauthors{Nicholls et al.}

\author{David C Nicholls\altaffilmark{1}, Michael A Dopita\altaffilmark{1,3,4}, Helmut Jerjen\altaffilmark{1}, Gerhardt R Meurer\altaffilmark{2}}

\email{\emaila}

\altaffiltext{1}{Mount Stromlo and Siding Spring Observatories, \\
Research School of Astronomy and Astrophysics, \\ Australian National
University, Cotter Road, \\ Weston Creek, ACT 2611, Australia. }
\altaffiltext{2}{International Centre for Radio Astronomy Research, The University of
Western Australia, M468, 35 Stirling Highway, Crawley WA, 6009 Australia}
\altaffiltext{3}{Astronomy Department, King Abdulaziz University, \\
P.O. Box 80203, Jeddah, Saudi Arabia}
\altaffiltext{4}{Institute for Astronomy, University of Hawaii, \\2680
Woodlawn Drive, Honolulu, HI 96822}

\begin{abstract}

Using an optically-unbiased selection process based on the HIPASS neutral hydrogen survey, we have selected a sample of 83 spatially isolated, gas-rich dwarf galaxies in the southern hemisphere with $cz$ between 350 and 1650 km s$^{-1}$, and with R-band luminosities and H~\textsc{i} masses less than that of the Small Magellanic Cloud. The sample is an important population of dwarf galaxies in the local Universe, all with ongoing star formation, and most of which have no existing spectroscopic data. We are measuring the chemical abundances of these galaxies, using the Integral Field Spectrograph on the ANU 2.3m telescope, the Wide Field Spectrograph (WiFeS). This paper describes our survey criteria and procedures, lists the survey sample, and reports on initial observations.
\end{abstract}


\section{Introduction}

The main epoch of galaxy formation, occurring at redshifts $1 \lesssim z \lesssim 4$ (e.g. \citeauthor{mad98} \citeyear{mad98}) is long past. Large galaxies have evolved far from the pristine state in which they formed. Mergers, galactic winds, cycles of star birth and death and outflows from active galactic nuclei have not only chemically enriched the interstellar medium of these galaxies, but have also chemically enriched their surrounding inter-galactic space \citep{kob07}. Theory predicts that chemical abundances will rise as newly-collapsing galaxies evolve and undergo successive generations of star formation. However, the spread of abundances should narrow with time as the interstellar gas becomes either internally mixed, diluted with infalling chemically pristine gas, or depleted by starburst-driven galactic-scale outflows \citep{del04,kob07}. Thus, the chemical abundance in galaxies is inextricably linked to star formation and galaxy growth, providing a record of the generations of cosmic star formation, mass accretion, and mass-loss in galaxies. 

In this regard, the gas-rich dwarf galaxies are of particular interest. These galaxies have processed a much smaller fraction of their gas through stars, and so are much less chemically evolved than the more massive disk or elliptical galaxies. Many of them have formed and developed in ``quiescent'' regions of space, possibly at very early times, away from the centres of dense clusters and the gravitational harassment of  massive neighbors, so their star formation and chemical evolution history may well be simpler than for the more massive systems.

A detailed study of the stellar and gas content of these gas-rich dwarf galaxies may therefore be expected to provide answers to the following key questions:
\begin{itemize} 
\item{What is the relationship between mass and chemical abundance for low mass galaxies? The answer could provide sensitive constraints on the mass fraction of the interstellar medium lost to inter-galactic space by galaxy winds. We would like to know whether matter lost from dwarf galaxies can account for the ``missing baryon'' problem---the discrepancy between the baryonic mass currently contained in visible galaxies, and the baryonic mass inferred from the standard model of cosmology \citep{bre07}.}

\item{What is the total mass of oxygen in each dwarf galaxy? This is determined
from the chemical abundance and the total H~\textsc{i} mass of these galaxies. Since we know how many
stars have formed in the galaxy from I-band photometry, we may obtain tight constraints on the fraction
of heavy elements that have been lost to the inter-galactic medium by galactic winds in these systems.}

\item{Is there a chemical abundance floor in the local Universe? Such a floor, predicted to be
about 1/100 solar \citep{kob07}, would result from pollution of inter-galactic gas by heavy elements ejected in
starburst-powered galactic winds, or by black hole jet-powered outflows from massive galaxies.}
\end{itemize}

There is evidence that dwarf glaxies have a wide variety of evolutionary histories (e.g. \citet{gre97,mat98,tol09}) and that---possibly as a result of this---they display a wide scatter on the mass-metallicity relationship \citep{tre04, lee06, gus09}. However, the observed sample of objects in this region of the parameter space is still relatively small. 

Although these dwarf systems are dominated by dark matter \citep{mat98} and they retain much of their original gas content, the inferred chemical yields are much lower than those estimated for more massive galaxies. This suggests that these galaxies may have formed very early in the history of the Universe and have subsequently had quite severe episodes of mass-loss through galactic winds. The study of small isolated gas-rich dwarf galaxies may provide evidence of the conditions and processes long since erased in larger galaxies and clusters.

Dwarf irregular galaxies (`dwarf' defined for the purposes of this study as having a gas+stellar mass less than that of the Small Magellanic Cloud) are numerous throughout the Local Volume but they remain surprisingly poorly studied. Optical catalogs by their nature tend to be incomplete and affected by the Malmquist Bias \citep{mal21}.  Dwarf galaxies can be missed due to their low surface brightness, or mistaken for much more massive and distant objects. The SINGG survey  \citep{meu06} went a long way towards rectifying this situation as it drew its objects from the HIPASS neutral hydrogen survey \citep{zwa04, mey04, kor04}. This provided a complete volume-limited sample for a given H~\textsc{i} mass content. Follow-up R-band and H$\alpha$ photometry revealed both the stellar luminosity and the star-formation rate in these galaxies. Much to the surprise of \citet{meu06}, evidence for on-going ($\lesssim 10$~Myr) star formation was found in nearly every case. This opens up the possibility of follow-up spectroscopy to establish the chemical abundances in the H~\textsc{ii} regions of these objects.

Going beyond the SINGG survey, which investigated only $\sim$ 10\% of the HIPASS sources, there are many other small isolated gas-rich dwarf galaxies in the HOPCAT catalogue \citep{doy05}, which could also provide insight into dwarf galaxy evolution. Both the SINGG and HOPCAT galaxies can now be studied very efficiently using integral field units (IFUs). In particular the Wide-Field Spectrograph (WiFeS; Dopita et al. 2007, 2010) at the Australian National University (ANU)
2.3m telescope at Siding Spring Observatory is ideally suited for the study of these galaxies. The WiFeS instrument is a highly efficient double-beam, image-slicing integral-field spectrograph with spectral resolutions $R=3000$ and 7000, covering the wavelength range 3500\AA~ to $\sim$9000\AA.  It offers a contiguous $25 \times 38$~arc sec field-of-view, well matched to the angular size of these objects. Such an instrument obviates the need to obtain optical photometry in either broad or narrow bands, and allows us to extract the complete spectra of each of the individual  H~\textsc{ii}  regions present in the galaxy. 

Motivated by the opportunity to study dwarf galaxy formation and evolution in the Local Volume (D $\lesssim$ 20 Mpc), we have identified a volume-limited sample of small, isolated gas-rich dwarfs. In this paper, we describe the characteristics of this sample, and describe initial results.

\section{The ``SIGRID'' sample}

It is conventional to identify dwarf galaxies using their optical characteristics (such as their morphological appearance or their spectrum). This has been done by surveys such as the Sloan and Byurakan \citep{yor00, aba03,mar67}.  Of necessity this introduces an optical Malmquist bias into the sample, whereby fainter galaxies are present in larger numbers close by, while brighter galaxies are more strongly represented at greater distances \citep{mal21, but05}.

We can avoid the optical bias problem---or at least substantially reduce it---by using the neutral hydrogen 21cm signatures of gas rich dwarf galaxies as a means of identifying the sample members. It is then necessary to identify optical counterparts of these H~\textsc{i} sources---H~\textsc{ii} regions and stellar populations. In the case of the HIPASS survey this was undertaken through the HOPCAT optical counterparts study which used COSMOS data as its optical source \citep{doy05}.  Taking this process further, \cite{meu06} presented a selection of 468 HIPASS objects for follow-up H$\alpha$ and R-band studies, of which to date 362 have been observed.  Using H~\textsc{i} surveys to identify galaxies was also used as a means of checking sample completeness in the 11HUGS survey \citep{ken08}. 

Both the SINGG and 11HUGS surveys include objects of a range of masses and luminosities, with and without close neighbors. In this work we target specifically small isolated dwarf galaxies.

Starting with the SINGG and HOPCAT catalogs, based on the HIPASS neutral hydrogen survey, we have identified a sample of 83 small isolated gas-rich dwarf galaxies in the Southern Hemisphere, in the Local Volume, beyond cz$\sim$350 km s$^{-1}$.  We have also drawn on the catalogs by \citet{kar08, kar11} for additional targets.

In addition to the SINGG data, the DSS and GALEX surveys have been used to identify objects that show no evidence of organised structure, but do show evidence of current star formation associated with the H~\textsc{i} and H~\textsc{ii} regions\footnote{It is worth noting that these sources trace somewhat different stellar populations: O and B stars in the GALEX UV objects and O stars in the SINGG H~\textsc{ii} regions.}.

Through this selection process we automatically exclude dwarf objects similar to LGS3 in the Local Group that retain some (``warm phase'') H~\textsc{i} but exhibit no current star formation \citep{you97,hun04}.  Using the reasoning from \citet{you97}, this implies the H~\textsc{ii}  regions in our objects arise from ``cold phase'' gas, with the inference that they represent cold gas inflow regions (whether or not these are a contributing source of star formation).  Objects with neutral hydrogen but lacking obvious UV and H$\alpha$ emission are therefore unlikely to have significant O- and B-star populations.

We have also excluded, as far as possible, low surface brightness (LSB) objects whose brighter regions might be mistaken for isolated dwarfs. An example of this is HIPASS J0019-22 (MCG-04-02-003), which from the GALEX images is clearly a faint face-on low surface brightness spiral galaxy, and identified as an LSB galaxy by \citet{war07}.  Likewise, we have used HIPASS velocity profile widths to exclude larger, side-on objects with substantial rotational velocities. We refer here to our sample as the Small Isolated Gas Rich Irregular Dwarf or ``SIGRID'' sample.

\section{Selection criteria}

The objects identified in this survey are gas-rich; they show evidence of current star formation; they are less luminous than the SMC; they have lower neutral hydrogen masses than the SMC; they are isolated; they are irregular or centrally condensed, with
no spiral structure evident; they are separate from major galaxy clusters; and they are generally located between the Local Group and the Fornax Cluster in distance.

We used the HIPASS \citep{mey04}, HOPCAT \citep{doy05} and SINGG \citep{meu06} catalogs and the NED and HyperLEDA databases to select objects having the following characteristics:

\begin{enumerate}[(i)]
  \setlength{\itemsep}{1pt}
\item {Gas rich with evidence of star formation (ionized hydrogen emission, bright in UV)}
\item {Low R-band absolute magnitude: \mbox{M$_{R}$ {$>$} -16.7}}
\item {Low neutral hydrogen mass:\mbox{ log$_{10}$(m$_{HI}$) {\textless} 8.7(M$_\odot$)} \footnote{Criterion (iii) was based on using the SMC as a yardstick, but it also emerged naturally from the sample selected using the other parameters: in the SINGG data, from which absolute R-band magnitudes could be determined reliably, only 11 objects meeting criterion (ii) had log$_{10}$(m$_{HI}/m_\odot$) $>$ 8.7. Of these, 8 were low surface brightness galaxies, one was in a congested field, and only two were potential SIGRID candidates, neither with exceptional log$_{10}$(m$_{HI}$).}}
\item{HIPASS H~\textsc{i} rotation velocity w$_{50}$ $<$ 130 km s$^{-1}$}
\item {Isolation: no immediate nearby neighbors or evidence of tidal effects}
\item {Located outside regions around nearby galaxy clusters where infall would distort redshift (see Table 2)}
\item {Irregular morphology showing no evidence of spiral structure (GALEX, DSS and SINGG)}
\item {Distance: redshift recession velocity between 350 and 1650 km s$^{-1}$ and flow corrected recession velocity $<$ 1650 km s$^{-1}$}
\item {Declination: from HIPASS, south of +2$^\circ$}
\end{enumerate}

Every object was inspected visually, using, where available, the SINGG H$\alpha$ and R-band images; the DSS optical images; and GALEX UV images.   Except where already observed in this program, the SIGRID sample only includes objects that show evidence from the SINGG or GALEX imaging of active star formation and young stellar populations. An example is given in Figure 1, which shows the DSS2, GALEX and SINGG composite color images of galaxy MCG-01-26-009 (HIPASS J1001-06; SIGRID 45).  These images demonstrate the optical appearance; the presence of H$\alpha$ emission; and the extended UV from a young stellar population.

There is some uncertainty as to whether this object is tidally influenced  by NGC 3115 at an angular separation of 91 arcminutes. There are several published semi-direct distance measurements for the larger object, including globular cluster and planetary nebula luminosity function methods. Taking the average of these measurements gives a distance that is somewhat larger than the flow-corrected redshift distance.  However, only the flow-corrected redshift distance is available for SIGRID 45, and while there may be systematic errors in the flow-corrected distances, they are likely to be similar for both.  As a result, using the flow-corrected distances in estimating tidal effects is probably the most reliable approach.  On this basis, SIGRID 45 is included in the sample.
\begin{figure*}[htpb]
\includegraphics[width=\hsize]{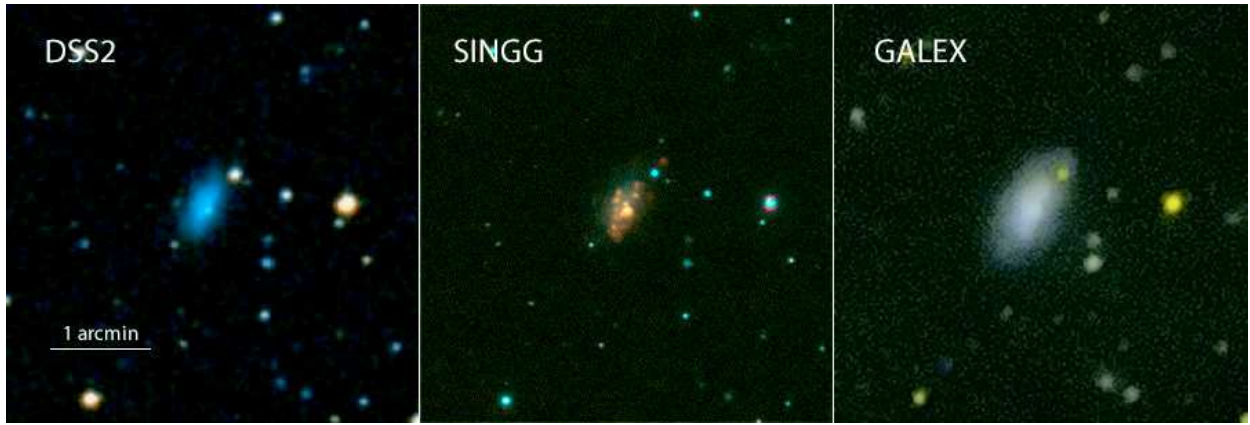} 
\caption{DSS2, SINGG and GALEX composite color images of galaxy MCG-01-26-009}
\end{figure*}
\FloatBarrier
The degree of isolation was estimated visually from these sources, and using the NED database. The isolation criterion was further refined the sample using variants of the ``main disturber'' techniques described by \citet{war07} and Karachentsev's tidal index \citep{kar04}, as described in \S 6.3 below.

From the 4500+ HIPASS objects, the 3600+ HOPCAT optical counterparts, the 462 SINGG objects and additional HIPASS identifications by \citet{kar08}, 83 galaxies have been selected using the above criteria and constitute the SIGRID sample.

Figure 2 shows a plot of the log neutral hydrogen mass, determined from the 21cm integrated intensity from the HIPASS catalog, vs the R-band absolute magnitude, determined from the HOPCAT R-band magnitudes and the HIPASS local heliocentric recession velocities. No attempt has been made in this graph to correct M$_R$ for local flows, as it is primarily intended to illustrate the sample size. The blue rectangle shows the SIGRID sample as a subset of HOPCAT. Note that not all the objects within the rectangle are included in the sample, as some have been excluded because of near neighbors, evidence of structure etc.

\begin{figure}[htpb]
\centering
\includegraphics[width=\hsize]{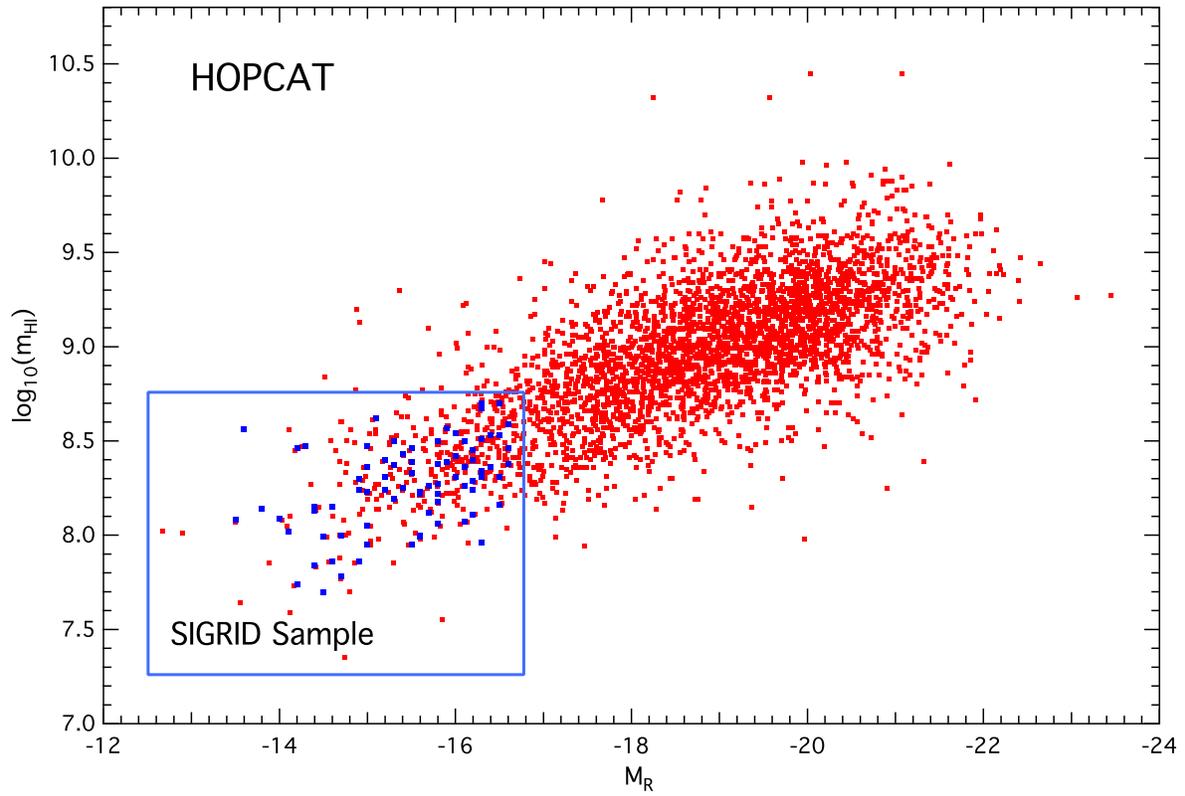} 
\caption{The HOPCAT catalog plotting log$_{10}$(m$_{HI}$) vs. M$_{R}$ showing the SIGRID sample region. Sample galaxies are shown in blue: some potential members have been excluded due to other selection criteria.}
\end{figure}

\FloatBarrier

\section{The SIGRID catalog}

Table \ref{table_1} shows the full SIGRID sample.  Velocity values are flow corrected using the \cite{mou00} model, including Virgo, Great Attractor and Shapley Supercluster infall, except where direct distance measurements are available.  These are converted to recession velocities taking H$_0$ = 73 km s$^{-1}$ Mpc$^{-1}$.  Neutral hydrogen masses are taken from the HIPASS catalog.  Absolute R-band magnitudes are derived from apparent magnitudes listed in the HOPCAT catalog, except where available from the SINGG catalog (52 of the 83 objects).  
\begin{figure*}[htpb]
\includegraphics[width=\hsize]{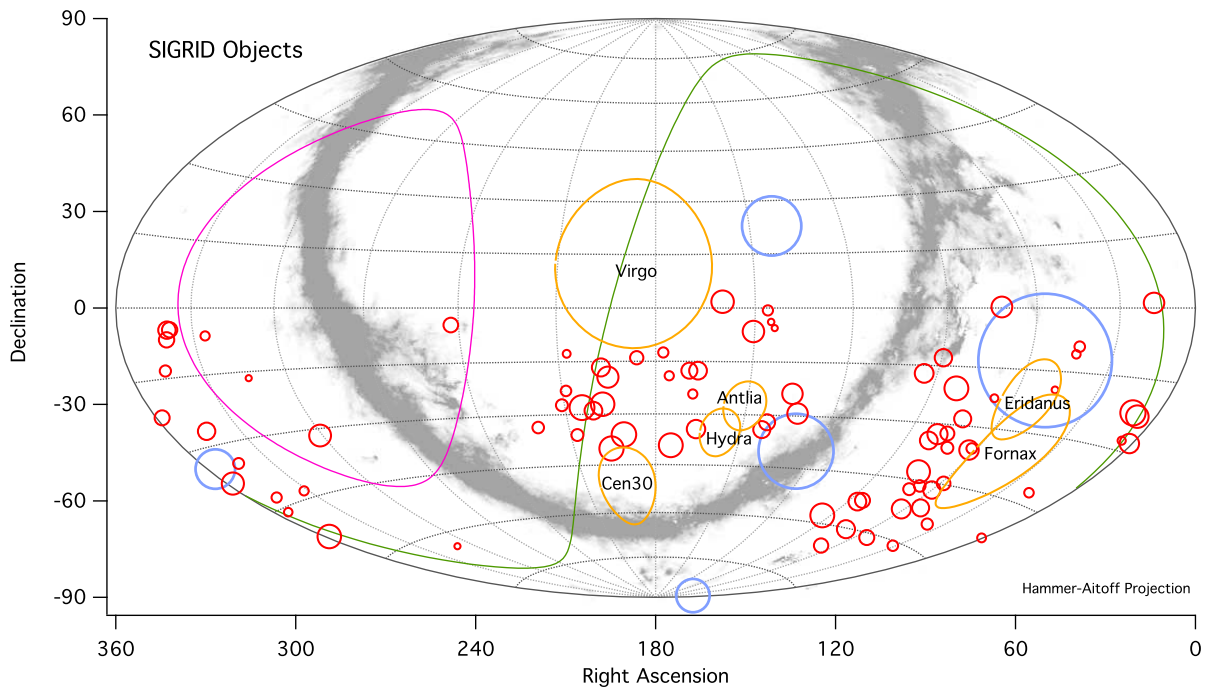} 
\caption{SIGRID galaxies with voids, galaxy clusters and MW dust absorption. SIGRID objects are plotted (red) with closer objects shown larger; the Local (Tully) Void is plotted with a radius of 50$\degree$ (magenta); Local mini-voids are shown as simple circles proportional to their size (blue) ; Galaxy clusters are plotted with radius values listed in Table 2 (yellow); MW dust absorption is from \citet{sch98}.}
\end{figure*}

Figure 3 shows the SIGRID objects plotted using a Hammer-Aitoff projection with the Local (Tully) Void, local mini-voids \citep{tik06} and galaxy clusters (Fornax, Eridanus, Antlia, Hydra, Virgo and Centaurus30).  Milky Way high dust absorption regions (gray) are calculated from \cite{sch98}.  The Supergalactic Plane is shown in green. 

The apparent lack of SIGRID candidates in the region below the Milky Way dust absorption region, between 300\degree  and 240\degree RA, is most probably due to the Local (Tully) Void  \citep{tul08,nas11}.  Only four objects in the SIGRID sample (\#68, 69, 70, 71) occur within the 50$\degree$ radius of this void, including the object [KK98]246 (SIGRID 68), currently the most isolated dwarf galaxy known and the only confirmed galaxy located within the void \citep{kre11}.\footnote{The authors are indebted to the reviewer for pointing out the nature of this object and the importance of its inclusion in the sample.}

It is evident in Figure 3 that none of the SIGRID sample are located near the centers of the main galaxy clusters. We have specifically excluded any galaxy that could be infalling into the cluster with high peculiar velocity, as discussed in $\S$6.

\FloatBarrier

\section{Sample images}

To illustrate the range of objects included in the SIGRID sample, we present here six images from the SINGG survey (figures 4 to 9).  These show the diversity of objects in the sample, from the active starburst region in Figure 5 to the faint H$\alpha$ regions in Figure 7; and the various distribution morphologies of the H~\textsc{ii}  regions.  

The H~\textsc{ii}  region morphologies of the objects in our sample may be classified as follows:

\begin{enumerate}[(i)]
 \setlength{\itemsep}{1pt}
\item{single or few H~\textsc{ii}  regions;}
\item{multiple H~\textsc{ii}  regions but centrally clumped;}
\item{multiple dispersed H~\textsc{ii}  regions; or} 
\item{separate H~\textsc{ii}  regions or dwarf galaxy pairs}
\end{enumerate}

Each object can be characterised in addition by the ratio of its H$\alpha$ and R-band fluxes, although this does not correlate exactly with the visual appearance in the SINGG images.

\begin{figure}[htpb!]
\includegraphics[width=\hsize]{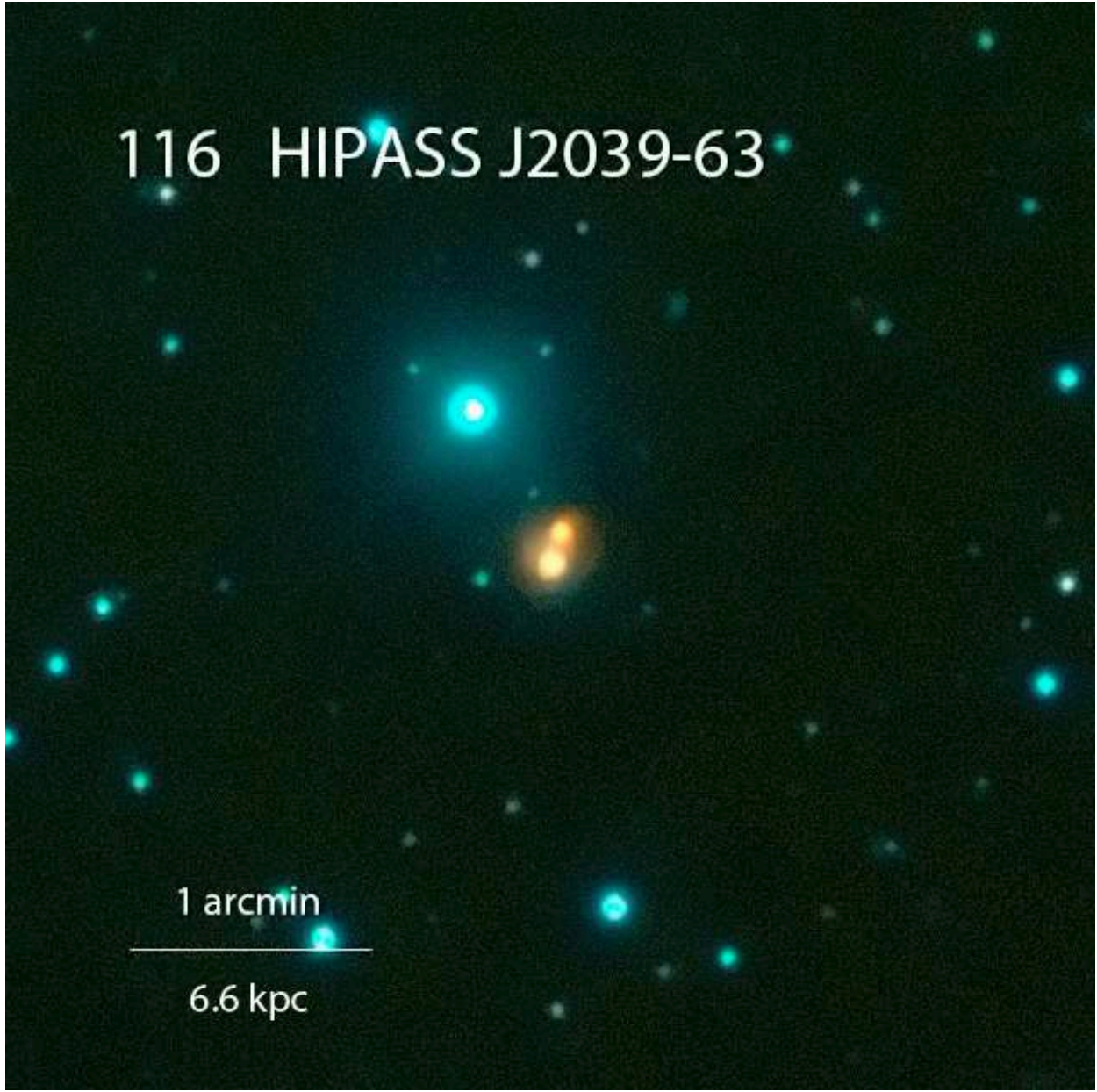} 
\caption{Type (i): Centrally condensed H$\alpha$ region(s)}
\end{figure}

\begin{figure}[htpb!]
\includegraphics[width=\hsize]{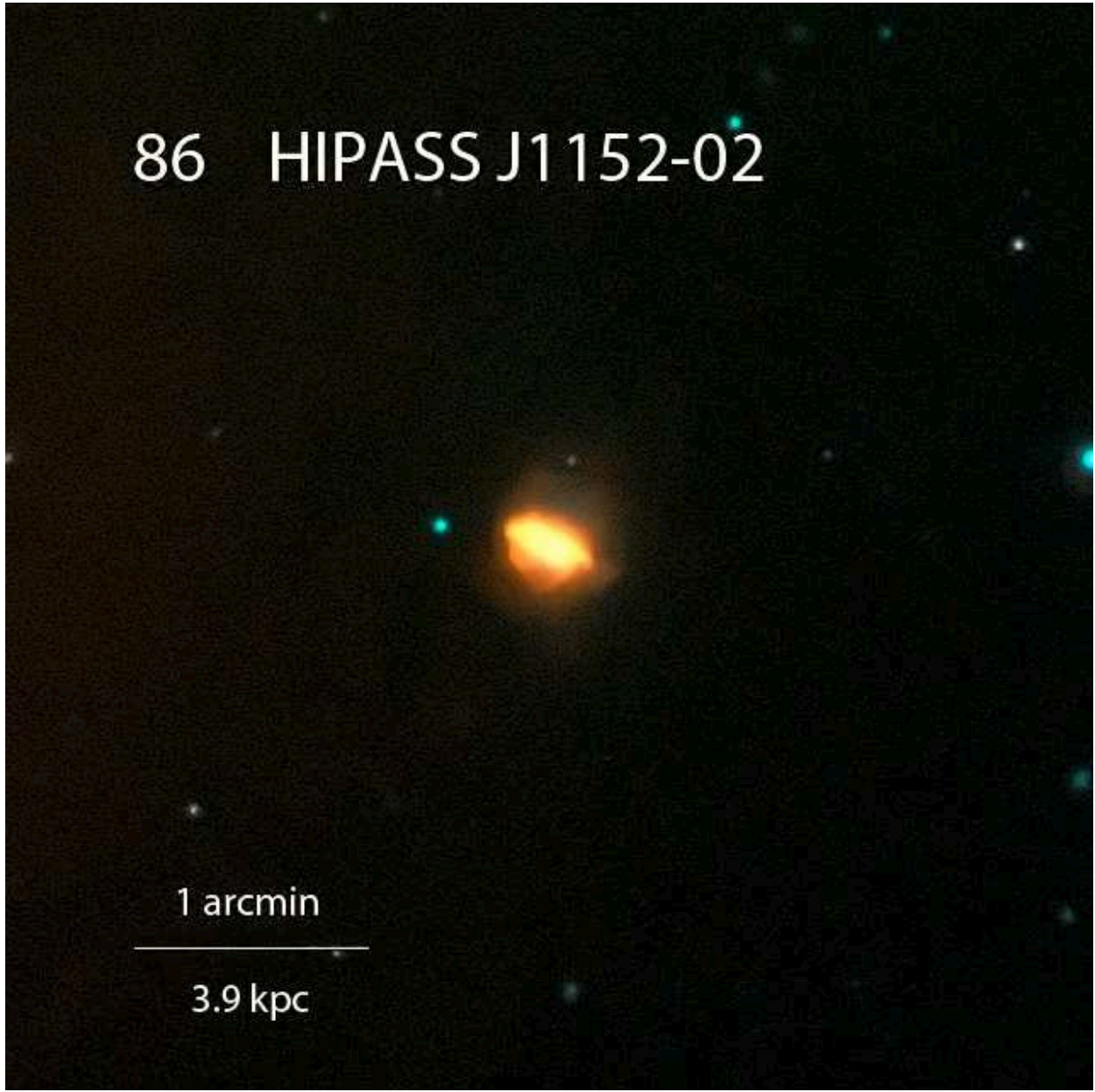} 
\caption{Type (i): Bright H$\alpha$ showing active starburst and ouflows}
\end{figure}

\begin{figure}[htpb!]
\includegraphics[width=\hsize]{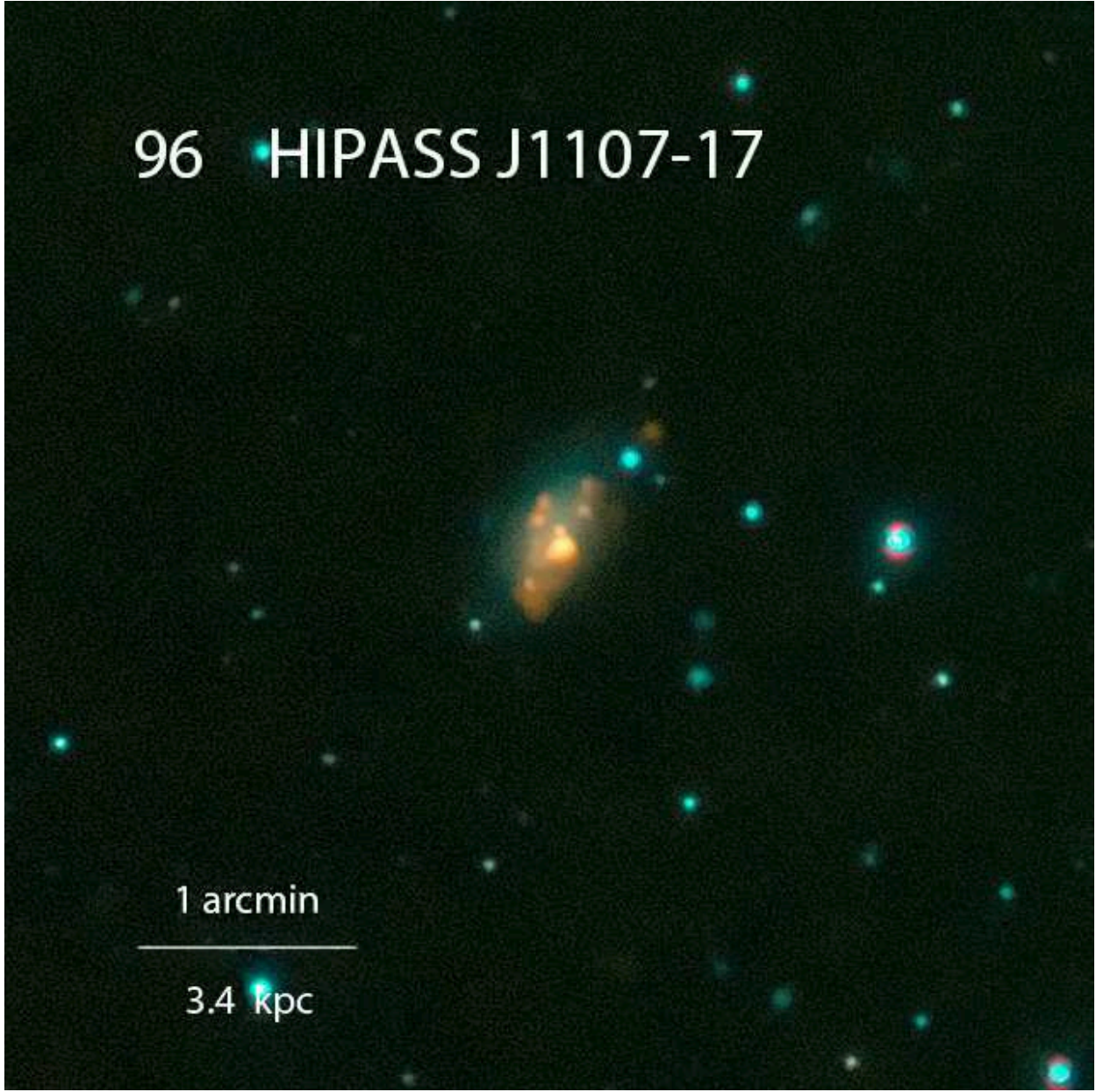} 
\caption{Type (ii): Centrally clumped multiple H$\alpha$ regions}
\end{figure}

\begin{figure}[htpb!]
\includegraphics[width=\hsize]{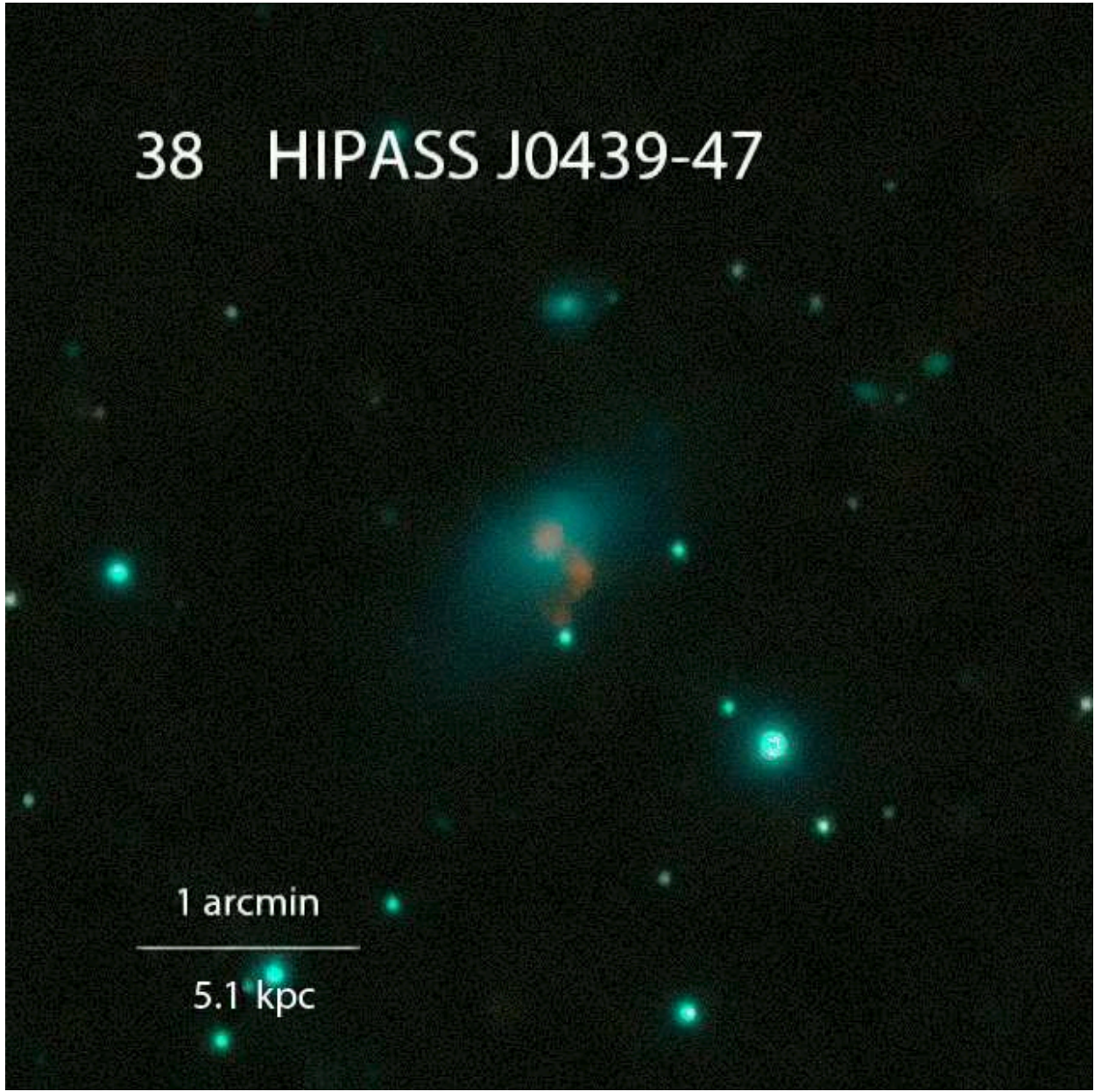} 
\caption{Type (ii): Faint H$\alpha$ compared to R-band}
\end{figure}

\begin{figure}[htpb!]
\includegraphics[width=\hsize]{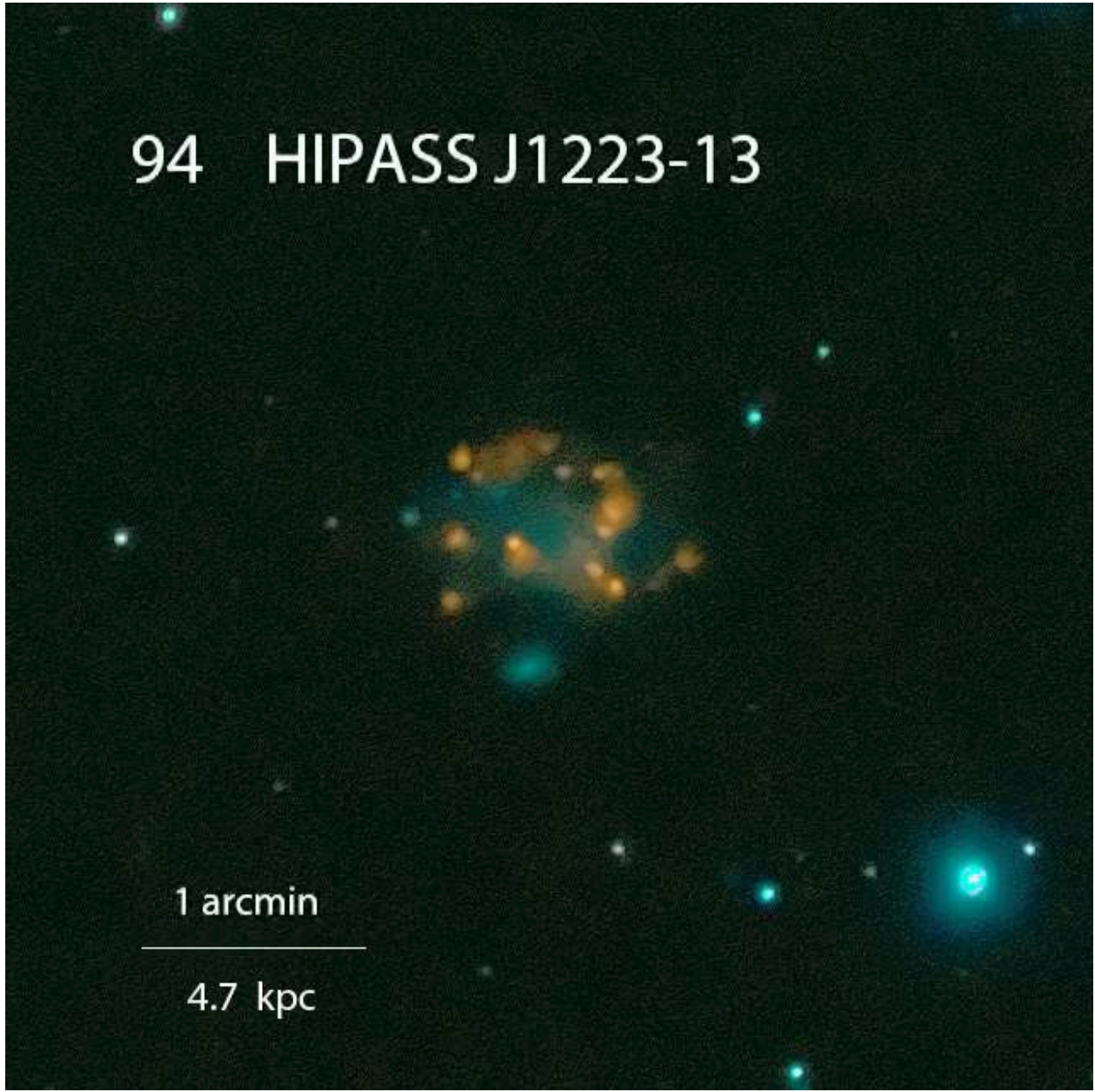} 
\caption{Type (iii): Dispersed H$\alpha$ regions}
\end{figure}

\begin{figure}[htpb!]
\includegraphics[width=\hsize]{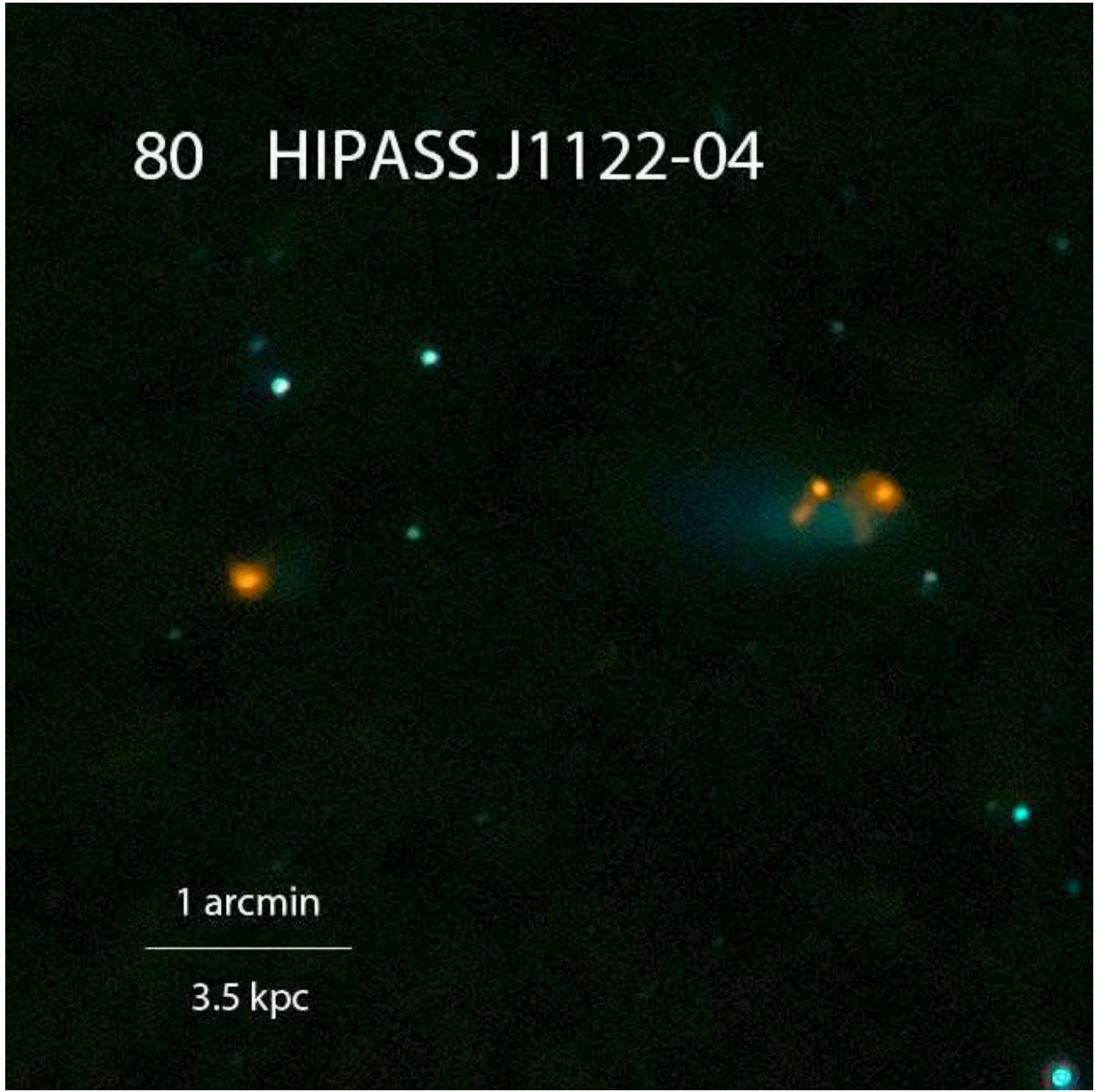} 
\caption{Type (iv): Separate neighboring H$\alpha$ regions}
\end{figure}

\FloatBarrier 

\section{Isolation}

A key aspect of the SIGRID sample is that it consists of isolated objects, implying that current star formation is intrinsic to the galaxy's environment and evolutionary processes, rather than triggered by tidal influences of nearby larger galaxies.  Isolation is also important because we wish to investigate galaxies that have formed from an intergalactic medium that is as far as possible pristine, or at least unenriched by outflows from large galaxies in recent times.

Thus there are two aspects of isolation we need to consider: isolation from nearby larger galaxies which could cause tidal effects; and isolation from galaxy clusters which might have enriched the IGM from which the dwarf galaxy formed.  We have tackled this in three ways: visual inspection of optical and ultraviolet survey images; calculation of tidal potentials arising from galaxies in the same general region; and identification of objects that might have large peculiar radial velocities arising from infall into and through clusters.

\subsection{Visual inspection of survey images}

The initial identification of isolated targets involved visual inspection of DSS and GALEX images to exclude objects obviously proximate to larger galaxies.  This process provided evidence of any nearby large neighbors within a few arc minutes, which was followed up using redshift data to estimate physical separation.  More distant influences were evaluated using estimates of association with galaxy clusters and calculation of tidal influences from larger galaxies beyond the range of the visual inspection. An example of where the visual process was important occurred with HIPASS J1158-19b (ESO572-G034).  While the initial tidal calculations did not suggest any significant effects, the DSS2 image showed it was within $\sim$16 arc minutes of  line-of-sight of two large galaxy/galaxy pairs, NGC 4027 and NGC 4038/9 (the Antennae galaxies).  The standard NED data for distance (on which the tidal calculations are based, $\S$6.3) suggest that the difference in redshifts (1114  and $\sim$1650 km s$^{-1}$)  and the fact that they are not part of a large cluster should take the small galaxy out of range the the larger ones.  However, there is evidence from TRGB distances that the larger galaxies are much closer, ie 13.3 Mpc ($\sim$970 km s$^{-1}$) \citep{sav08}.  The closer distance implies a significantly greater prospect for tidal effects, and could explain the strong starburst and outflows apparent in the SINGG image of the dwarf galaxy.

\subsection{Excluding candidate galaxies due to proximity to galaxy clusters}

When identifying possible candidates for the SIGRID sample we need to consider whether objects close to the same line-of-sight as more distant galaxy clusters (Centaurus, Eridanus, Fornax etc.) are in fact foreground objects, or whether they are members of the cluster with large peculiar velocities due to infall. Using recession velocity as a proxy for distance to determine the degree of isolation and absolute magnitude does not work in cluster environments, due to our inability to distinguish between normal Hubble flow recession and peculiar velocities due to cluster infall. We may therefore need to exclude from the sample objects located in the direction of nearby clusters.  The size of the `exclusion zone' depends on the redshift of the cluster center and size parameters such as the virial and zero-infall (turnaround) radii of the cluster and  the velocity dispersion within the cluster. 

\cite{kar10} investigated blueshifted galaxies in the Virgo Cluster, which they explain in terms of high inflall peculiar velocities of objects located beyond the cluster center. They found that these objects are confined to line-of-sight locations within the projected virial radius of the cluster (6\degree).  They also found that 80\% of such objects have young populations (S, BCD or dIr).  They interpreted this as indicating these galaxies are still in the process of falling into the cluster, from beyond the cluster.

This suggests that galaxies with young stellar populations in the same line-of-sight as a galaxy cluster might be infalling from the far side of the cluster. This is important in the SIGRID context as all members have young stellar populations. Even if they exhibit apparent redshifts much lower than the cluster average, and thus appear to be closer than the cluster, it is not possible to say conclusively that they are not distant infalling objects, blueshifted against the Hubble flow trend.  An example of this may be galaxy CCC026, identified as a  member of the Cen30 Cluster \citep{jer97}.  Its heliocentric recession velocity is 1438 km s$^{-1}$.  The recession velocity of the Cen30 cluster is 3397$\pm$139 km s$^{-1}$ and its velocity dispersion $\sigma$ is 933$\pm$118 km s$^{-1}$ \citep{ste97}.  If CCC026 has a peculiar velocity $\sim$ 2$\sigma$ this object may be a cluster member infalling from beyond the cluster center, on which basis it has been excluded from the SIGRID sample.

The distortion of the flow velocity is well illustrated in calculations by \cite{ton00} (their Figure 1) which show an `s-curve' velocity variation with distance for Virgo Cluster infalling galaxies. The non-monotonic behavior leads to three possible distance solutions for objects in direct line-of-sight with a cluster, but the effect falls off with increasing angular separation from the cluster center.  Near the center of the cluster the effect is capable of reducing substantially the apparent recession velocity/redshift distance of infalling cluster members. At the 2$\sigma$ level, the model shows that objects with apparent heliocentric recession velocities as low as 250 km s$^{-1}$ can readily be found near the Virgo cluster center,  falling in from beyond the cluster.  The blue-shifted galaxies reported by \cite{kar10} are extreme cases of this phenomenon. Such infalling galaxies are thus much more distant than if their redshifts were due solely to Hubble expansion.  This effect was also explored by \citet{tul84} for the Virgo Cluster, who showed that objects as far as 26$\degree$ from the cluster center can exhibit anomalous redshifts (their Figure 4).  In these circumstances, redshift cannot be used as a distance measure for estimating isolation or absolute magnitudes. 

The Virgo Cluster also affects the SIGRID sample membership despite its northern location.  While the cluster center lies over 10$\degree$ from the nearest potential SIGRID object, its zero infall radius (taken here as 25$\degree$ based on several literature values) extends well into the southern hemisphere.  On this basis, 14 candidate objects have been excluded from the sample.  All objects in the final sample lie at or outside the projected zero infall velocity radius for the cluster \citep{kar10a}, implying that they are not bound by the cluster's gravitational potential. 

The question of potential infall into a cluster is also important for objects in line-of-sight with the Centaurus30, Fornax, Antlia and Eridanus clusters.  To exclude rogue galaxies with high peculiar velocities, we have adopted a selection criterion to exclude any sample candidate that lies within the circle of the projected zero-infall (turnaround) radius of a nearby cluster, where the candidate's recession velocity is within $\pm3\sigma$ of the cluster recession velocity.  Information on the zero-infall radius for all the relevant clusters is not available from published information, but we have adopted the values shown in Table \ref{table_2}.  We have attempted to err on the side of caution.  We have also confirmed that the objects are not associated with any of the Southern Compact Groups \citep{iov02}.  

We have further considered whether candidates are outlying members of smaller association of galaxies---galaxy groups or sheets---and whether this should rule them out as sample members. As there are numbers of gas-rich dwarf galaxies in the Local Group with distances from the major galaxies $>$ 270 kpc---implying they have had little opportunity to lose their gas through interaction with larger group members---\citep{grc09}, in general we have not excluded from the sample galaxies that may be associated with the outer regions of groups and sheets.

Because the virial and zero-infal radii are less well defined for groups and associations (e.g. the NGC5128/NGC5236 group), they are not a good guide to interactions.  In this case, we need to calculate the tidal influences to indicate likely present or past interactions with larger galaxies.  To explore this we have used a `disturber index' (\S 6.3 below) to check whether the tidal effects of adjacent galaxy group members are significant.

\subsection{Tidal indices}
Lack of apparent optical correlation of SIGRID objects with larger neighbors does not rule out the existence of potential tidal disturbers outside the fields from DSS and GALEX that could otherwise be identified visually.  Thus it is important to identify potential disturber galaxies by calculating the tidal effect of galaxies in the same general region as the targets.  We have investigated several ways of calculating this.

It should be noted that the purpose of calculating a disturbance index is not to obtain precise values for tidal forces, which in many cases is impossible, but to flag potential sample members which may have been tidally influenced in the past.  As a result, a precise value for tidal strength is not required, and approximate methods can be used.

\cite{km99} (also \citeauthor{kar04} \citeyear{kar04}) developed the `tidal index' $\Theta$ as a way of estimating the isolation of a galaxy from the effects of tidal disturbance.  Their work looked at nearby galaxies with heliocentric recession velocity V$_h$ $<$ 720 km s$^{-1}$, where the distances to the target galaxies and to potential tidal disturbers are known through direct measurement.  They used estimates of the masses of the galaxies based on their distances and H~\textsc{i} rotational velocities, allowing evaluation of an expression of the form
\begin{equation}
\Theta = log(M\times d^{-3}) +C\nonumber
\end{equation}
where $M$ is the disturber galaxy mass, $d$ is the separation between disturber and target and $C$ is an arbitrary constant, evaluated by setting the `cyclic Keplerian period' to the Hubble time, a point beyond which galaxies could be deemed not to have interacted \citep{km99}.  For each target, $\Theta$ was taken as the maximum value of the index for all disturbers.

For the SIGRID galaxies, many of which are at greater distances (V$_h$ up tp $\sim$1650 km s$^{-1}$) than the Karachentsev catalog, the set of disturbers identified using the NASA/IPAC Extragalactic Database (NED) is incomplete due to optical bias effects.  Most SIGRID distances are greater than measurable by direct methods (for example, measuring the Cepheid or RR Lyrae variable stars, or estimating distances using the tip of the red giant branch) and can usually only be estimated from redshifts or the Tully-Fisher relation. More important, H~\textsc{i} rotation curves are not available for many of the potential disturbing galaxies, making a direct measurement of mass difficult or impossible. This implies that $\Theta$ cannot be calculated reliably, or at all, for the SIGRID galaxies.   As a result, we have explored other disturbance measures.

There are a number of possible measures of interaction between target and disturber.  The simplest is to calculate a hypothetical `time of last contact', based on the distance between disturber and target and assuming a velocity of separation of 100 km s$^{-1}$ ($\sim$100kpc/Gyr).  We have also taken a more sophisticated approach, based on the Karachentsev tidal index, to develop a simple alternative tidal index to estimate isolation for the galaxies in the SIGRID sample.

We define a `disturber index', $\Delta$, as:
\begin{equation}
\Delta =C_1\times (log_{10}(\frac{ L_{B}}{d_{Mpc}^3}) -C_2)\nonumber
\end{equation}
where \textit{L$_{B}$} is the disturber absolute luminosity (B band) and $d_{Mpc}$ is the separation between the disturber and target galaxy in Mpc. \textit{C$_1$} and  \textit{C$_2$} are constants (see below).  $\Delta$ is computed from NED data for all disturbers within the 10 degree `near name lookup' NED limit from the target, and within the range of recession velocities  $V_h(target)$ $\pm$250 km s$^{-1}$.  Typically this search generates a list of between 1 and 50 potential disturber galaxies for a target galaxy, for each of which a value of $\Delta$ is calculated.  As with the $\Theta$ index of Karachentsev et al., the disturber index for a target galaxy is the maximum of the individual tidal potentials for all the possible disturbers identified by NED.  This index has the great virtue that the necessary information on magnitudes and heliocentric recession velocities is readily available. 

\textit{L$_{B}$} is used as a proxy for disturber galaxy mass, so $\Delta$ is in effect a measure of the tidal potential at the target galaxy due to the disturber, following the approach described by \cite{km99}.  The R- or I-band magnitudes might provide a better stellar mass estimate, but these are not available for the majority of NED listings.  2MASS magnitudes would provide even better mass estimates but are not available for most objects.

The distances from the observer to the target and disturber galaxies were initially calculated assuming $V_h$ is a measure of actual distance due to Hubble expansion and contains no peculiar velocity or local flow component. Angular separation and the V$_h$ of the target galaxy are used to calculate lateral separation. The total separation for each galaxy pair is then calculated using Pythagoras.  Thus $\Delta$ is a proxy for $\Theta$ and can be calculated from NED data.  

The constants \textit{C$_1$} and  \textit{C$_2$} are evaluated  by deriving a linear best fit for $\Delta$ to $\Theta$ for 16 galaxies common to the Karachentsev and an earlier extended SIGRID list. The values \textit{C$_1$} = 0.894 and  \textit{C$_2$}  = 13.0 have been adopted. They give values for $\Delta$ which indicates isolation when $\Delta$ $<$ 0. Table \ref{table_3} compares values of $\Theta$ and $\Delta$ calculated for the galaxies common to SIGRID and the Karachentsev et al. catalog (2004). Objects not part of the final SIGRID list are included in Table 3 to allow better comparison between the two methods.  They were excluded from the SIGRID list for reasons such as cluster line-of-sight proximity discussed earlier (\S6.2).

While the relationship between $\Theta$ and $\Delta$ is not completely consistent, $\Delta$ does appear to be a reasonable substitute for $\Theta$. Both the $\Theta$ and $\Delta$ indices identify the possibility of disturbance of the target galaxy by the disturber, but because we do not know, for most objects, their actual locations and velocities, a positive value of the index does not guarantee that there has been interaction.  The methods are approximate, but negative values of $\Theta$ and $\Delta$  are a good indication of isolation, and this is their purpose.  The `time of last contact' can be used to clarify marginal cases.

It should also be noted that this approach is limited by the available 10$\degree$ radius of available NED data.  As a result, potential disturbers outside this radius will be missed.  However, the ready availability of reasonably consistent and complete data makes this a useful technique.  The more analytically based technique of \citet{km99} is itself limited by the accuracy of available distance measurements, and may be best suited to objects closer than V$_h$$\sim$720 km s$^{-1}$.

\subsection{Absolute magnitude bias for low luminosity objects}

In the lists of possible disturbers generated for each target galaxy, the majority are usually faint objects, with a few larger (typically NGC catalog) objects.  We need to take care to estimate the masses of the fainter objects carefully to identify those that, while small, may be close enough to disturb the target galaxy tidally.

As \cite{mat98} has shown, the King formalism in which mass follows light leads to an underestimate of mass for low luminosity objects (L \simlt 10$^8$\Lsolar). Adopting the finding by  \cite{str08} that there is a common mass scale that applies to dwarf Milky Way satellites,  indicating a minimum integrated mass (dark + baryonic)  of 10$^7$ \Msolar within 300pc of galactic center for galaxies with luminosities $<$ 10$^8$ \Lsolar, we have found that taking into account very low luminosity objects that are relatively close to our targets makes little difference to the estimated maximum tidal potentials experienced by SIGRID objects, as the potentials are dominated by larger more distant galaxies. 

\subsection{Dust reddening corrections to distance and luminosity}

We have corrected SIGRID apparent magnitudes for Milky Way dust absorption using the dust maps from \cite{sch98}.  NED also provides E(B-V) values for these objects using the same dust maps.  At the time of this work, there were slight discrepancies between NED values and values calculated directly using current Schlegel et al. code, but they do not materially influence the results.

\subsection{Local flow corrections to distance and luminosity}

Bearing in mind that precise values of the disturber index are not required, how accurately must distances be known to estimate magnitudes suitable for use in the disturber index?  In practice the potential disturbers are experiencing similar flow fields to the SIGRID targets, so it is reasonable to use the apparent recession velocities V$_h$ of both as the basis for approximate absolute magnitude calculations in the disturbance index. This has the virtue that the values are readily available using the NED `near name' search. Using flow-corrected velocities we  recalculated the tidal indices for a 10\% subset of SIGRID objects.  It makes no significant difference to the isolation calculations whether we use the flow corrected distances or the simple V$_h$ distances, confirming our assumption.

The flow model  takes into account corrections to the measured heliocentric recession velocities, V$_h$, arising from the motion of the Sun around the Galaxy, our motion towards the Local Group barycenter, and flows induced by mass concentrations in the local Universe, such as the Virgo Cluster, the Great Attractor and the Shapley Supercluster.

Flow models investigated (following the approach used by \cite{meu06} for the SINGG survey), were those published by \cite{mou00} and \cite{ton00}.  The Mould et al. model turns out to give the better results when compared to directly measured distances.

Of the mass concentrations that could influence recession velocities, the Virgo Cluster at 17 Mpc \citep{jer04} appears to have the greatest effect.  This is not surprising because of the distance limit imposed on the SIGRID sample, V$_h$ $<$ 1650 km s$^{-1}$. However, in estimating absolute magnitudes, we have used the flow corrections for Virgo, Great Attractor and Shapley Supercluster infall as these give the best fit to direct distance measurements (where available).  Twenty-one of the SIGRID sample have such distances listed. We have also used the \textit{Extragalactic Distance Database} \citep{tul09} as an additional source of independently measured distances.  The closer objects have tip of the red giant branch measurements and some of the more distant objects have distances estimated using the Tully-Fisher relation.  We conclude that the flow corrected velocities give reliable distance values for use in absolute magnitude calculations and that any peculiar velocity effects are minor, but as recommended in the NED database documentation, we have used the direct distance measurements when available instead of the flow model values in calculating absolute magnitudes.

We have also considered how to deal with the effects of anomalous velocities of galaxies away from voids, explored by \citet{tul08} and \citet{nas11}. Understanding the details of the flow away from such voids and the uncertainties in the size and location of voids, makes a complete understanding of local flow patterns extremely complex.  As our aim in this work is to select a sample of galaxies that are isolated, we have assumed the simpler flow-model approach as the principle means of avoiding disturbed objects.

We conclude that the SIGRID galaxies have not experienced significant tidal influences from larger adjacent galaxies within at least the last 5 Gyr---in many cases, ever---and therefore that their recent star formation episodes are intrinsic to the galaxies themselves.

\section{Sample completeness}

It is worth noting that completeness in the normal sense is not essential to the SIGRID sample, but isolation is important. Our purpose is to obtain a sample for spectroscopic study that can be reasonably argued to be isolated. By setting stringent sample selection criteria, we automatically exclude numbers of dwarf galaxies that would be present in a full set of Local Volume dwarf galaxies. However, it is of interest to consider how complete the base set is from which the SIGRID sample has been selected.

The completeness of the set of gas-rich dwarf galaxies in the Local Volume identified in this way is determined by the completeness of the HIPASS survey, i.e. 95\% or better for an integrated 21cm flux of 9.4 Jy.km s$^{-1}$ \citep{zwa04}. Thus the set of dwarf galaxies is largely complete for galaxies of the SMC neutral hydrogen mass to a distance of $\sim$1250 km s$^{-1}$. Beyond this distance for SMC-size objects, and closer for smaller galaxies, there does exist a bias, but this is measurable and understood. Figure 10 shows M$_R$ plotted against distance for the SIGRID galaxies.  An unbiased sample would show a horizontal line. The top of the distribution is reasonably flat due in part to sample cut-off, but the base shows some bias, arising from the H~\textsc{i} detection limit of the HIPASS catalog.

\begin{figure}[htpb]
\includegraphics[width=\hsize]{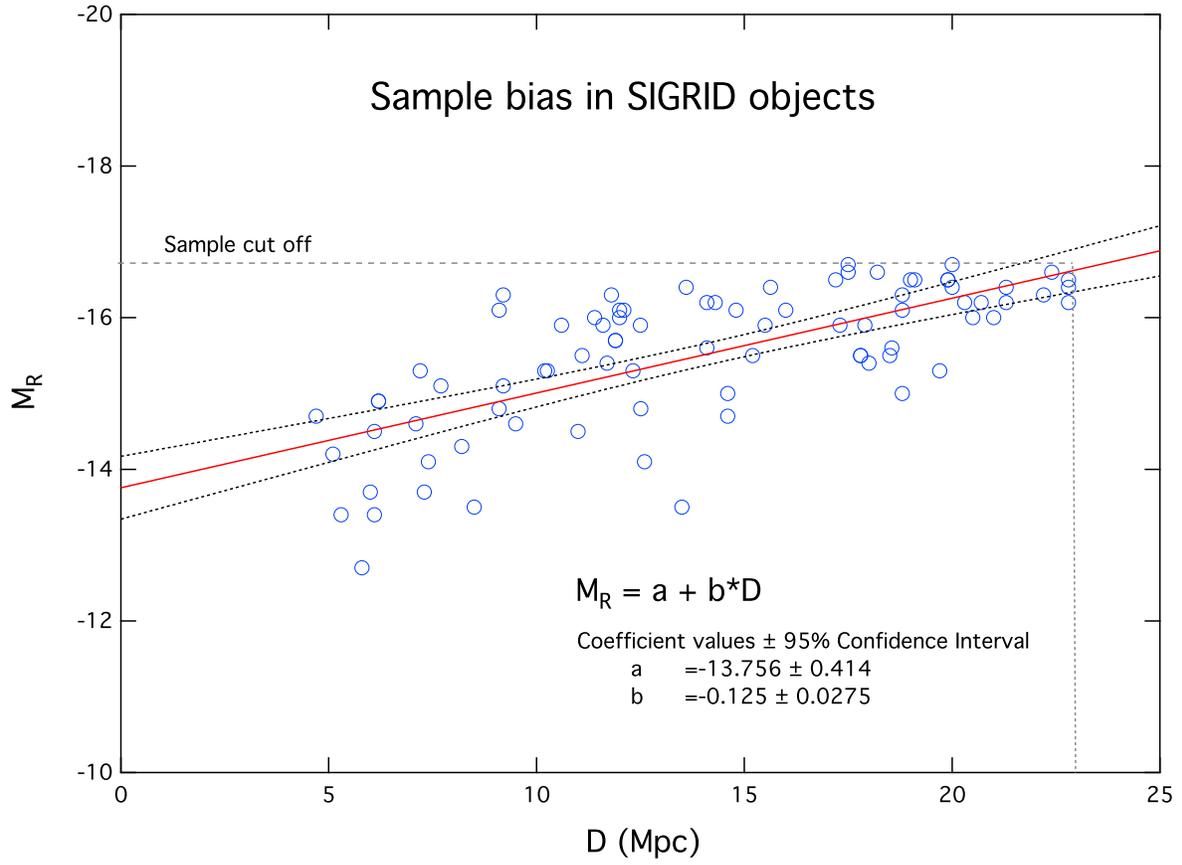} 
\caption{Sample bias for the SIGRID objects} 
\end{figure}

\begin{figure}[htpb]
\includegraphics[width=\hsize]{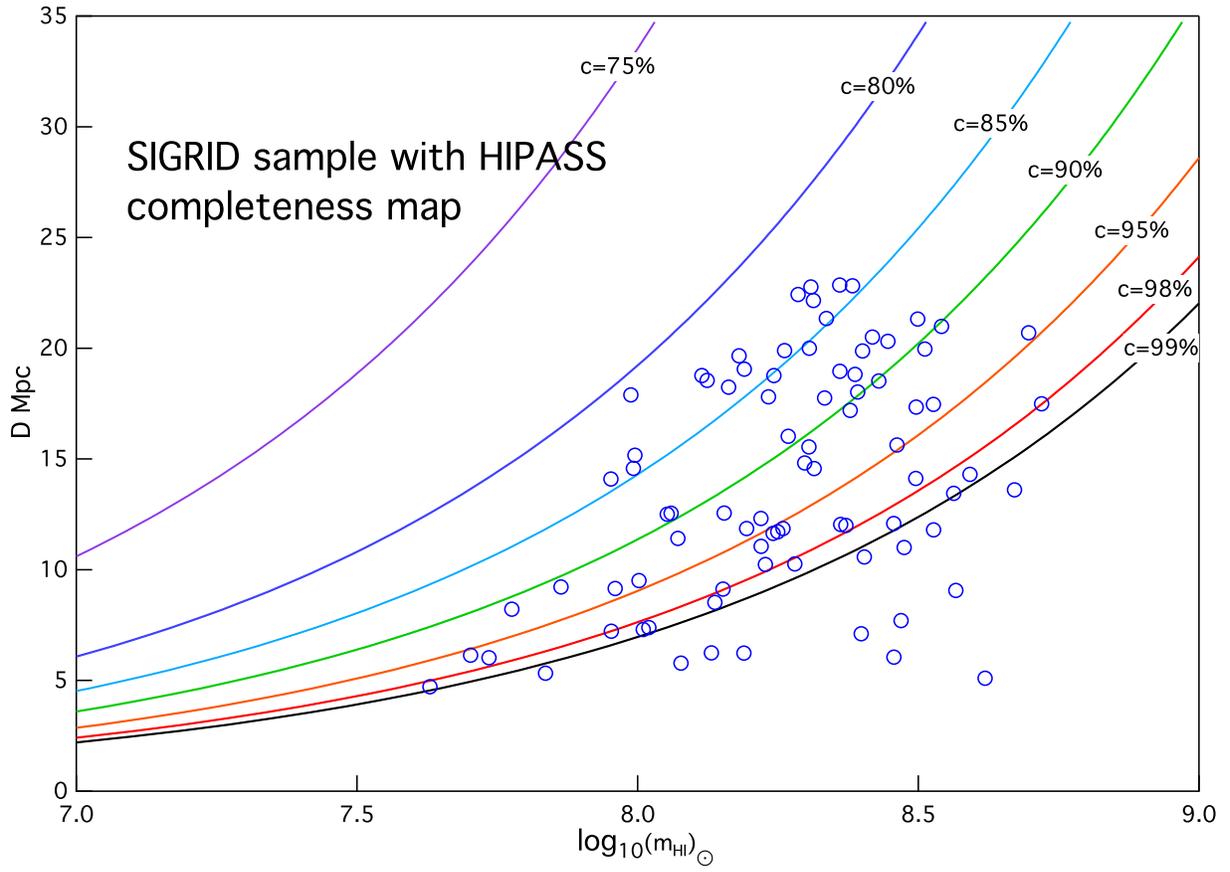} 
\caption{SIGRID sample plotted over HIPASS completeness contours} 
\end{figure}

Figure 11 shows the HIPASS sample completeness contours computed from a formula derived from that given by \citet{zwa04}:
\begin{equation}
log_{10}(m_{HI}) = log_{10}(\frac{erf^{-1}(C)}{0.12} - 6.4) + 5.371 + log_{10}(D^{2})\nonumber
\end{equation}
\noindent where D is the distance in Mpc and C is the sample completeness.  Over 40\% of the SIGRID 
sample lies above the HIPASS 95\% completeness level.

\section{Metallicity}

Metallicity, the abundance of elements heavier than Helium, is an important parameter that controls many aspects of the formation and evolution of stars and galaxies \citep{kun00}. Observations show that larger galaxies have higher metallicities than smaller galaxies, the so-called ``mass-metallicity relation''. A major aim in our measurements of nebular metallicities is to improve understanding of the mass-metallicity relation for low mass, low luminosity isolated dwarf galaxies.

The mass-metallicity relation has been well studied for higher mass galaxies, but low luminosity objects have received much less attention \citep{tre04, lee06, gus09}. There has so far been no complete or systematic study of the metallicity distribution in low mass isolated gas rich dwarf galaxies in the Local Volume.   

In this work we have identified a number of previously unstudied low mass, low luminosity objects, where we are using a single observational base to increase consistency and reduce scatter.  The main focus of this work is to measure the metallicities of this sample, using strong line and, where possible, direct metallicity techniques \citep{kew02, kew08}. In all our measurements to date (see \S9 below), excellent S/N data have been obtained on the standard nebular emission lines.

In addition, in a few objects, a number of  He \textsc{i} and He \textsc{ii}  lines have been detected. Intense OH airglow lines make observing longer wavelength faint nebular lines difficult in these objects, as integration times for the objects are typically 1 to 2 hours.  The time variability of both the intensity and, particularly, the relative rotational level populations in the atmospheric OH molecules \citep{nic72} makes the removal of such interference problematic, although the `nod and shuffle' technique used by the WiFeS spectrograph works very well.

Gas-rich dwarf galaxies characteristically show substantially lower nebular metallicities than larger galaxies, implying a much lower chemical yield from any previous star formation episodes. The low metallicities may be explained in several ways:

First, the star-formation chronology \citep{lee09, mcq09, mcq10}: 
\begin{itemize}
\item{Star formation only early, mainly prior to reionisation;}
\item{Continuous star formation at a low rate since the formation of the galaxy;}
\item{Occasional, irregular short bursts of star formation, fed by cold inflows of near-pristine gas; or}
\item{Only relatively recent commencement of star formation.}
\end{itemize}

Second, evolutionary factors (e.g. \citeauthor{kun00} \citeyear{kun00}):
\begin{itemize}
\item{Low retention of enrichment by supernova outflows due to shallow gravitational potential and low efficiency of retention of outflows  within the local H \textsc{ii} region;}
\item{Dilution of nebular metallicity by pristine gas inflows; and}
\item{Lack of interactions, mergers, tidal effects or active galactic nuclei to stimulate strong star burst and consequent enrichment.}
\end{itemize}

It is plausible that all of these have contributed to the chemical enrichment in the galaxies in our sample, and we anticipate that our measurements will help clarify the relative importance of these processes. \cite{dis08} have suggested that galaxy evolution may be simpler than it appears, but have not been able to identify the ``controlling parameter'' that describes the process.

We do, however, tend to favor the idea that, after some initial star formation before or around the re-inoisation era, later star forming episodes in dwarf galaxies have been sporadic, with durations of perhaps 0.5Gyr, resulting from inflows of cold pristine H~\textsc{i} (i.e. ``cold phase'' gas), whose frequency depends on the availability of such inflows.  This would imply that more isolated galaxies, arising in regions with a scarcity of gas suitable for cold inflows and initial galaxy formation, should show lower metallicity on average than those in more densely populated regions, where the IGM would have had more initial enrichment, and where the cold inflows would have been more frequent.  Although not stated by \citet{pus07}, this is an implication that could be drawn from their observations of blue compact galaxies in local voids.

SIGRID sample objects exhibit H~\textsc{ii}  emission, implying current star formation with a timeframe of $\sim$5 Myr, and clear evidence of UV emission from GALEX observations, implying a significant population of O and B stars, and consequent star formation over $\sim$50 Myr.  There is also evidence in initial observations with WiFeS (\S9 below) in a few objects of Balmer absorption lines in the associated stellar continuum, implying a robust intermediate age A star population, evidence of star formation for $\sim$400 Myr, similar to that found for NGC 839 \citep{ric10}.  

This is fully consistent with the findings of \cite{mcq09} that dwarf gas rich galaxies show evidence of star formation over periods of 200-400 Myr.  As those authors have also found, this demonstrates that star formation in a dwarf galaxy can occur for protracted periods of at least 0.4 Gyr, and that star formation can occur over all time scales in that period, ruling out global ``self-quenching'' of starbursts over shorter periods \citep{mcq09, mcq10}.

\section{WiFeS and nebular abundance measurements}

The WiFeS spectrograph is a new double beam image slicing IFS, designed specifically to maximize throughput from the ANU 2.3m telescope at Siding Spring. It covers the spectral range 320 to 950nm, at resolutions of 3000 and 7000. It has a science field of view of 25x38 arc seconds \citep{dop07}.

As many of the SIGRID objects subtend angles less than its FOV, WiFeS is an ideal instrument to measure nebular metallicities in the ionized hydrogen star forming regions. The instrument generates a data cube which allows exploration of nebular and continuum spectra in different regions of the target objects. Typically even in poor seeing WiFeS resolves SIGRID object star formation regions easily, making possible exploration of excitation and abundances in different regions of each object.

\begin{figure}[htpb!]
\includegraphics[width=\hsize]{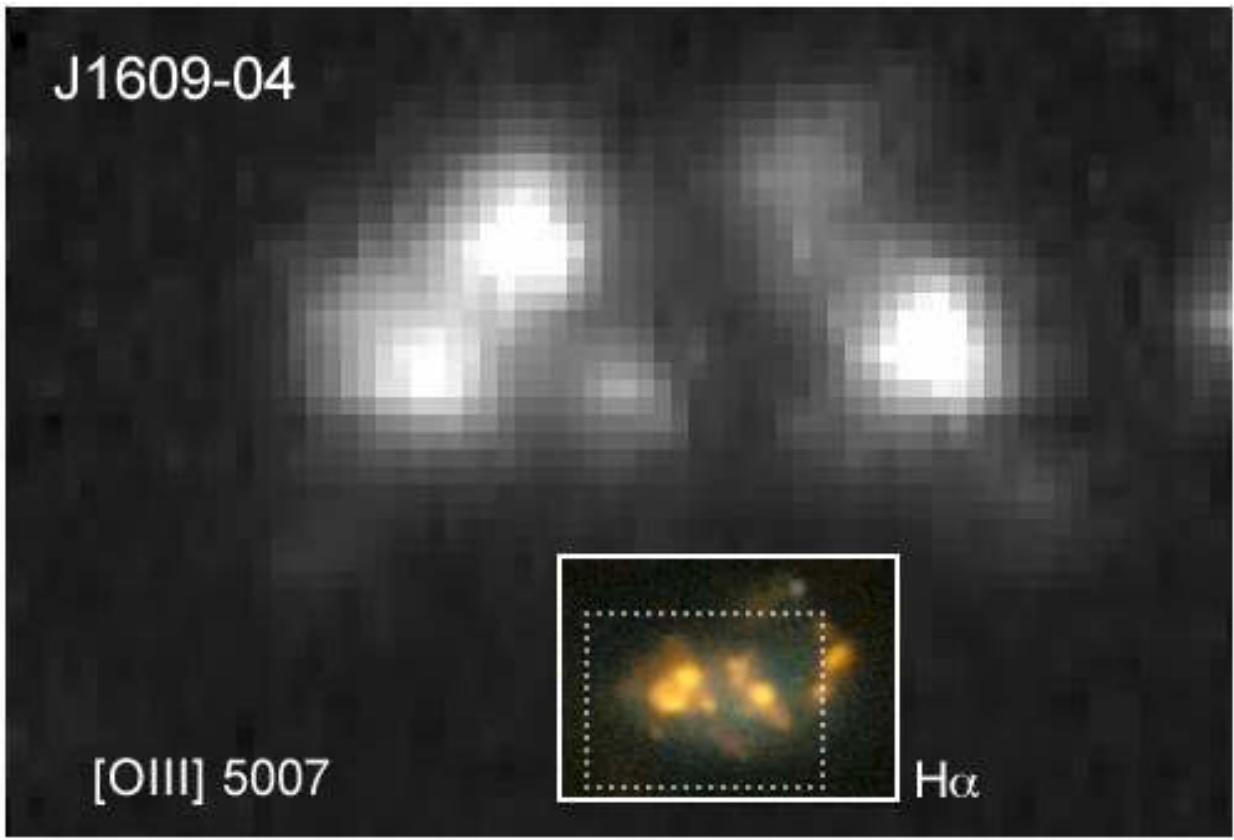} 
\caption{WiFeS datacube image of HIPASS J1609-04 at [\textsc{O iii}] 5007\AA, with SINGG image inset, showing WiFeS aperture} 
\end{figure}

Figure 12 shows a slice through the data cube of object HIPASS J1609-04 in [\textsc{O iii}] 5007\AA. The star forming regions are typically 5 arc seconds across, corresponding to a diameter of $\sim$300pc at 13.5 Mpc.

Inset in the figure is a composite image from the SINGG observations covering a slightly larger area, showing the H$\alpha$ regions orange and the stellar continuum (R-band) as cyan \citep{meu06}.

At this scale, the data cube allows spectra to be obtained for complete H~\textsc{ii}  regions around individual star-forming areas in a single observation. This will eliminate the bias toward high excitation regions at the centres of HII regions which are present in single-slit observations, and will allow for the derivation of more accurate chemical abundances from either the strong-line technique, or the direct electron temperature based techniques.  A further benefit is that photometry can be undertaken using an image from an appropriately weighted wavelength slice.

An additional benefit of IFU spectroscopy is that it avoids fiber-size sampling errors that can occur with surveys such as the Sloan Digital Sky Survey, where fiber diameters have a 3 arcsecond size on the sky  \citep{yor00}, and may miss important regions within a complex object such as illustrated in Figures  6, 8, and 9.

\section{`Sub-dwarf' galaxies?}
When we undertook visual inspection of the DSS and GALEX image fields, some potential SIGRID objects appeared to be associated with one or more adjacent small UV-bright star forming regions. Few if any of these apparent star forming regions have been catalogued, so it is not possible without additional spectroscopy to determine whether they are associated with the SIGRID objects, or random line-of-sight associations with more distant objects. However, consistent with this hypothesis, several SINGG objects do appear to have companions with active star formation regions---see for example, Figure 9.

If the objects are physically associated, it is possible that some of them may be extended loosely associated star-forming regions in otherwise extremely faint dwarf galaxies.  \cite{war07} concluded that there was a minimum number of stars a galaxy could form based on its initial baryonic mass, but as there appears to be no lower limit on initial masses, there should be no reason to impose a lower limit on the size of dwarf galaxies. If a power law describes the sizes of newly formed galaxies, one might expect regions with localised star-forming regions, where the IGM has condensed into numbers of very small `sub-dwarf' galaxies. In isolated regions, such low luminosity objects would have been catalogued only by chance, consistent with the objects discussed here. 

\cite{wer10} have identified what they term ``ELdots''---emission line dots---small isolated regions of H~\textsc{ii}  emission near galaxy-centered H~\textsc{ii}  sources. However, the isolated star forming regions reported here (should they be such) are unlikely to be the same phenomenon as ELdots, as the latter are unresolved sources associated with highly disturbed regions.

We have not excluded from the SIGRID sample potential members adjacent to these regions, where the objects are clearly isolated from other sources of tidal disturbance. We propose to explore such regions further, as they may cast light on the formation of galaxies at the smallest scales.

\section{`Bloaters'}

Another phenomenon we have found during visual inspection of SIGRID candidates we term `bloaters'---dwarf irregular galaxies with typically low absolute R-band magnitude and low H~\textsc{i} mass, but which appear to have a much larger physical extent than expected (typically 10 kpc or greater).  These objects give the appearance of having been disrupted by tidal interaction, but since they are isolated, as far as can be determined, recent gas inflow may explain them. Even if these objects were intially larger galaxies that have undergone strong starburst and have ejected most of their gas, they would still be intriguing objects.

An example is shown in Figure 13 from DSS1 imagery. This shows galaxy UGCA051 (HIPASS J0315-19; SIGRID 18).  Calculating its distance from its flow-corrected recession velocity as 22.2 Mpc (assuming H$_0$ = 73 km s$^{-1}$ Mpc $^{-1}$) gives a maximum linear dimension of $\sim$16 kpc, which is much larger than one would expect for a normal dwarf galaxy of this luminosity and neutral hydrogen mass  (M$_R$ = -16.3; log$_{10}$(mH~\textsc{i}) = 8.3$(M_\odot)$).  The possibility of it being a faint side-on LSB galaxy is unlikely, given the low HIPASS rotational velocity measurement w$_{50}$ = 48.7 km s$^{-1}$. It has been classified variously as dIrr, IB(s)m and LC V-VI.

There are several cases of `bloaters' in the SIGRID sample, but it will be necessary to undertake quantitative isophotal evaluation of the images to obtain reliable estimates of their size, to validate the visual obesrvations. Whether one considers these as single larger faint objects, or as a small group of merging dwarf galaxies, may be a matter of interpretation. Certainly they are much more extended than, for example, 1Zw18 \citep{van98}.  We are tempted to speculate that objects such as [KK98]246 (SIGRID 68) which show extended HI regions around a much smaller stellar core \citep{kre11} may be the precursors of these bloated objects.

\begin{figure}[htpb]
\includegraphics[width=\hsize]{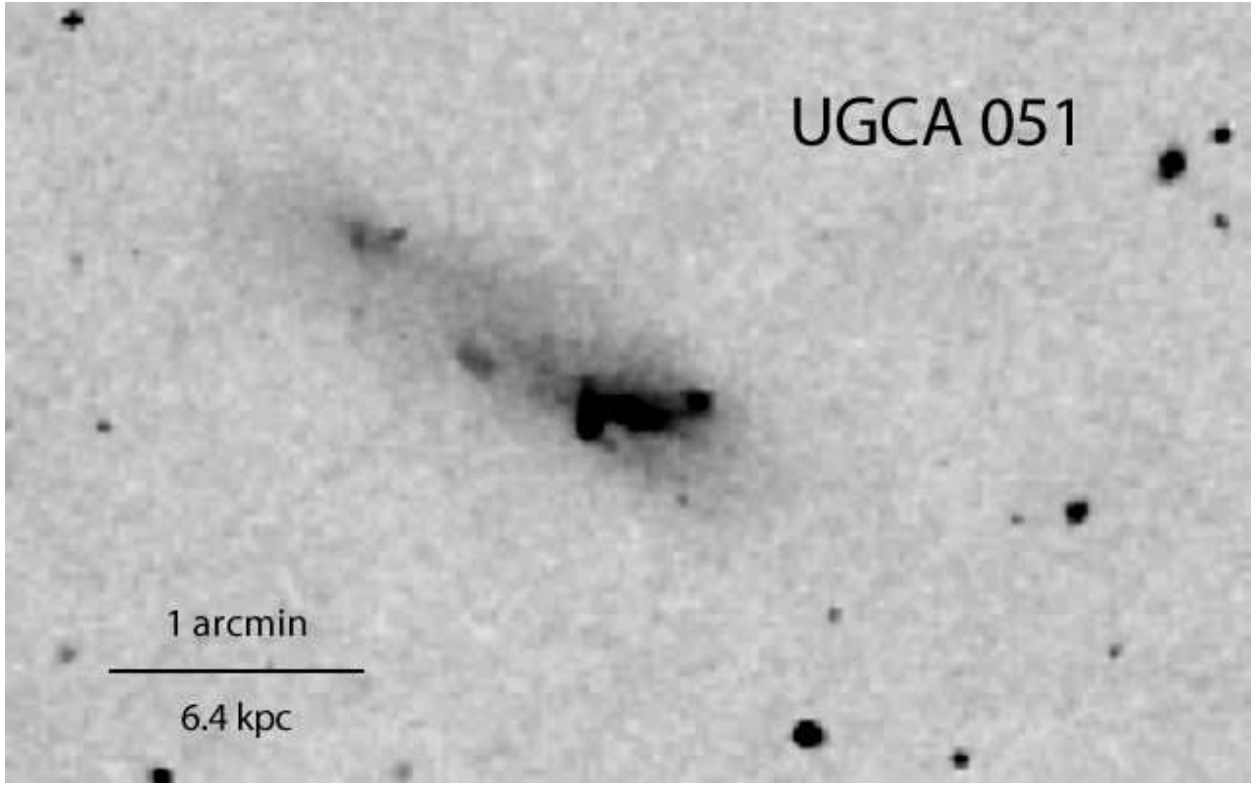} 
\caption{Example `bloater': a dwarf irregular galaxy with absolute R-band magnitude = -16.3, log(H~\textsc{i}) mass = 8.31$(M_{\odot})$, D = 22.2 Mpc, which appears to have a much larger physical extent than expected. Its H~\textsc{i} rotational velocity w$_{50}$ = 48.7 km s$^{-1}$ suggests it is unlikely to be a large side-on galaxy.} 
\end{figure}

\FloatBarrier

\section{A Blue Compact Dwarf excess?}

Blue Compact Dwarf (BCD) galaxies are an important class of dwarf galaxy exhibiting current active star formation. They have attracted considerable attention over the past 30 years as they harbor examples with the lowest nebular metallicity yet recorded and also provide information on the level of primordial helium \citep{izo04, izo07}. However, BCDs are not common among galaxy surveys (e.g. \citeauthor{izo04a} \citeyear{izo04a}) and are believed to be rare in groups and clusters \citep{cel07}.

First identified by \citet{sar70}, BCDs have been categorised in several ways, without clear consensus on consistent parameters \citep[e.g.]{loo86, sun02}. The simplest definition is that BCDs are ``dwarf irregular galaxies whose optical presence is exemplified by active region(s) of star formation'' \citep{sun02}. To this should be added the requirement that the galaxy and its star formation areas be centrally condensed rather than dispersed (i.e. compact).  To formalize the qualitative description, \cite{gil03} used a quantitative approach to the classification of BCD galaxies based on stellar mass, peak surface brightness and color.  However, the formal definition of ``BCD'' remains to be settled.

Due to the previously accepted scarcity of BCDs, it came as a surprise that 23\% of the dwarf galaxies identified in the SIGRID sample  in our initial visual investigation could be described as BCDs.  Examples of two possible BCD candidates for which SINGG images are available are shown in Figures  4 and 5.

To take these visual evaluations further, a more quantitative approach such as that used by \cite{gil03} will be necessary. We also propose to refine the `BCD yield' estimate by measuring `compactness' analytically, e.g. as per \citet{cel07}.  Although it is probable that some of these objects may not warrant the BCD classification after more careful measurement, our provisional conclusion is that using optically blind neutral hydrogen sampling methods may provide an efficient method for finding BCD galaxies.

\section{Conclusion}

We present a sample of 83 small isolated gas-rich dwarf irregular galaxies identified using their neutral hydrogen 21cm signatures and presnece of star formation.  The sample consists of galaxies with lower neutral hydrogen masses and lower R-band luminosities than the SMC. They are located in the southern sky with heliocentric recession velocities between 350 and 1650 km s$^{-1}$.  We are observing these objects using the WiFeS integral field spectrograph on the ANU 2.3m telescope at Siding Spring, to measure nebular metallicities.  We intend to use this data  to explore the mass-metallicity relation at the low mass end of the spectrum, and to see if there is evidence for a metallicty floor in the IGM. The sample appears to include a higher percentage of Blue Compact Dwarf Galaxies than expected from other optical surveys, and a number of unusual extended dwarf objects. \\ \\

\begin{acknowledgments}
Nicholls \& Dopita acknowledge the financial support of the Australian Research Council (ARC) through Discovery  project DP0984657.  This research has made use of the NASA/IPAC Extragalactic Database (NED) which is operated by the Jet Propulsion Laboratory, California Institute of Technology, under contract with the National Aeronautics and Space Administration. We acknowledge the use of the HyperLeda database (http://leda.univ-lyon1.fr). We acknowledge the comments and suggestions of the anonymous referee.
\end{acknowledgments}

\newpage

\begin{deluxetable}{rllrrrcrcrrr}
\rotate
\centering
\tabletypesize{\tiny}
\tablecaption{Small Isolated Gas-Rich Irregular Dwarf Galaxy sample, SIGRID\label{table_1}}
\tablewidth{625pt}
\tablehead{
\colhead{SIGRID} & \colhead{HIPASS} & \colhead{Optical galaxy} & \colhead{RA} & \colhead{\hskip -0.2in (J2000) \quad Dec} &
\colhead{V$_{hel}$} & \colhead{D} & \colhead{w$_{50}$} & \colhead{log$(m_{HI}$)} & \colhead{M$_R$} & \colhead{Tidal} & \colhead{Type}  \\[-1ex]
\colhead{ID} & \colhead{ID} & \colhead{ID} & \colhead{(h m s)} & \colhead{\hskip 0.15in(d m s)} &
\colhead{(km s$^{-1}$)} & \colhead{(Mpc)} & \colhead{(km s$^{-1}$)} & \colhead{$log(M\odot)$}  & \colhead{(mag)} & \colhead{index} & \colhead{} 
}
\startdata
1 & J0002-52 & ESO149-G013$\dagger$ & 00 02 46.3 & -52 46 18 & 1500 & 20.3 & 111.4 & 8.45 & -16.2 & -1.4 &  \\ 
2 & J0005-28 & ESO409-IG015 & 00 05 31.8 & -28 05 53 & 726 & 10.2 & 52.8 & 8.23 & -15.3 & -2.1 &  \\ 
3 & J0023-27 & 6dFJ0023042-275537 & 00 23 04.2 & -27 55 37 & 1539 & 21.0 & 121 & 8.54 & -16.0 & -0.5 &  \\ 
4 & J0031-22 & ESO473-G024* & 00 31 22.5 & -22 45 57 & 550 & 7.3 & 47.9 & 8.01 & -13.7 & -0.9 & iii \\ 
5 & J0043-22 & IC1574*$\dagger$ & 00 43 01.8 & -22 13 34 & 363 & 4.7** & 43.9 & 7.63 & -14.7 & -1.0 & iii \\ 
6 & J0107+01 & UGC00695 & 01 07 46.4 & 01 03 49 & 628 & 9.2 & 59.6 & 7.86 & -15.1 & -2.8 &  \\ 
7 & J0110-42 & ESO243-G050 & 01 10 48.8 & -42 22 31 & 1472 & 19.7 & 60.6 & 8.18 & -15.3 & -2.3 &  \\ 
8 & J0206-60 & ESO114-G028 & 02 06 15.8 & -60 56 24 & 1451 & 18.8 & 68.8 & 8.11 & -16.1 & -1.7 &  \\ 
9 & J0231-54 & 2dFGRS S857Z501 & 02 31 55.2 & -54 33 06 & 1394 & 18.0 & 85.3 & 8.39 & -15.4 & -1.3 &  \\ 
10 & J0253-09 & SDSSJ025328.63-085905.5 & 02 53 28.6 & -08 59 06 & 1423 & 18.8 & 87.1 & 8.24 & -15.0 & -0.1 &  \\ 
11 & J0255-10 & APMUKS(BJ) B025254.79-110123.4 & 02 55 19.6 & -10 49 17 & 1575 & 20.7 & 99.7 & 8.70 & -16.2 & -1.1 &  \\ 
12 & J0305-19 & UGCA051 & 03 05 58.7 & -19 23 29 & 1672 & 22.2 & 48.8 & 8.31 & -16.3 & -0.1 &  \\ 
13 & J0334-51 & ESO200-G045* & 03 35 02.2 & -51 27 13 & 1030 & 12.51 & 47.2 & 8.05 & -14.8 & -0.5 & iii \\ 
14 & J0334-61 & AM0333-611 & 03 34 15.3 & -61 05 48 & 1171 & 14.6 & 97.9 & 8.31 & -15.0 & -1.3 &  \\ 
15 & J0354-44 & ESO249-G027 & 03 54 29.3 & -44 45 13 & 1227 & 15.63 & 101.8 & 8.46 & -16.4 & -0.4 &  \\ 
16 & J0359-46 & AM0358-465 & 03 59 56.3 & -46 47 06 & 1018 & 12.32 & 80.8 & 8.22 & -15.3 & -0.3 &  \\ 
17 & J0406-52 & NGC1522 & 04 06 07.9 & -52 40 06 & 907 & 10.6 & 102.4 & 8.40 & -15.9 & -1.1 &  \\ 
18 & J0408-35 & ESO359-G022 & 04 08 45.6 & -35 23 22 & 1429 & 18.55 & 85.3 & 8.12 & -15.6 & -0.6 &  \\ 
19 & J0411-35 & ESO359-G024* & 04 10 57.5 & -35 49 52 & 850 & 10.26 & 69.7 & 8.28 & -15.3 & -1.4 & ii \\ 
20 & J0427-22 & ESO484-G019 & 04 27 19.9 & -22 33 34 & 1632 & 21.3 & 75.3 & 8.34 & -16.2 & -1.1 &  \\ 
21 & J0428-46 & ESO251-G003 & 04 28 41.2 & -46 19 16 & 1391 & 17.8 & 65 & 8.23 & -15.5 & -1.2 &  \\ 
22 & J0434-65 & AM0433-654 & 04 33 54.7 & -65 41 52 & 1239 & 15.2 & 38.5 & 8.00 & -15.5 & -0.9 &  \\ 
23 & J0439-47 & ESO202-IG048* & 04 39 49.2 & -47 31 41 & 1368 & 17.5 & 57.9 & 8.53 & -16.7 & -1.2 & ii \\ 
24 & J0446-35 & ESO361-G009 & 04 46 57.9 & -35 54 54 & 1348 & 17.3 & 97.3 & 8.50 & -15.9 & -1.2 &  \\ 
25 & J0448-60 & ESO119-G005 & 04 48 17.1 & -60 17 38 & 989 & 11.6 & 74 & 8.24 & -15.9 & -0.8 &  \\ 
26 & J0448+00 & UGC03174 & 04 48 34.5 & +00 14 30 & 669 & 9.1 & 105.6 & 8.57 & -16.1 & -0.7 & ii \\ 
27 & J0455-28 & APMUKS(BJ)B045339.97-282253.5* & 04 55 39.1 & -28 18 11 & 998 & 12.6 & 62.4 & 8.15 & -14.1 & -1.3 & iii \\ 
28 & J0457-42 & ESO252-IG001*$\dagger$ & 04 56 58.7 & -42 48 14 & 660 & 7.1 & 67.4 & 8.40 & -14.6 & -2.0 &  \\ 
29 & J0503-32 & ESO422-G025 & 05 03 45.9 & -32 19 51 & 1215 & 15.5 & 100.1 & 8.31 & -15.9 & -1.1 &  \\ 
30 & J0517-32 & 6dFJ0517216-324535* & 05 17 21.6 & -32 45 35 & 796 & 9.5 & 65.3 & 8.00 & -14.6 & -0.5 & ii \\ 
31 & J0523-34 & AM0521-343*$\dagger$ & 05 23 23.7 & -34 34 29 & 960 & 11.9 & 68.4 & 8.19 & -15.7 & -0.4 & iii \\ 
32 & J0527-20 & ESO553-G046 & 05 27 05.7 & -20 40 41 & 541 & 6.1 & 62.3 & 7.70 & -14.5 & -2.2 &  \\ 
33 & J0536-52 & ESO204-G022 & 05 36 26.0 & -52 11 03 & 1292 & 14.3** & 85.4 & 8.59 & -16.2 & -1.5 &  \\ 
34 & J0543-52 & ESO159-G025 & 05 43 06.2 & -52 42 02 & 1101 & 11.8** & 97.8 & 8.53 & -16.3 & -0.9 &  \\ 
35 & J0558-12 & LCSBL0289O & 05 58 02.3 & -12 55 47 & 911 & 12.0 & 101.3 & 8.36 & -16.1 & -0.5 &  \\ 
36 & J0615-57 & ESO121-G020 & 06 15 54.2 & -57 43 32 & 578 & 6.1** & 68 & 8.46 & -13.4 & -1.9 &  \\ 
37 & J0617-17 & HIPASS J0617-17$\dagger$ & 06 17 53.9 & -17 09 04 & 855 & 11.0 & 42.4 & 8.47 & -14.5 & -1.6 &  \\ 
38 & J0848-26 & ESO496-G010 & 08 49 06.0 & -26 19 18 & 809 & 9.2 & 61 & 7.96 & -16.3 & -1.2 &  \\ 
39 & J0903-23 & ESO497-G004 & 09 03 03.1 & -23 48 31 & 806 & 9.1 & 95.2 & 8.15 & -14.8 & -0.8 &  \\ 
40 & J0927-32 & UGCA165 & 09 27 25.9 & -32 00 35 & 1086 & 13.6** & 83.5 & 8.67 & -16.4 & 0.0 &  \\ 
41 & J0931-34 & ESO373-G006 & 09 31 50.0 & -34 08 17 & 1046 & 12.5 & 66.1 & 8.06 & -15.9 & -0.2 &  \\ 
42 & J0935-05 & 6dFJ0935505-053441* & 09 35 50.5 & -05 34 41 & 1499 & 22.8 & 49.6 & 8.36 & -16.4 & -1.9 & iii \\ 
43 & J0940-03 & [RC3]0938.0-0340* & 09 40 25.9 & -03 53 07 & 1453 & 22.4 & 53.6 & 8.29 & -16.6 & -1.5 & iii+iv \\ 
44 & J0944-00b & SDSSJ094446.23-004118.2 & 09 44 43.7 & -00 40 20 & 1223 & 18.8 & 124.8 & 8.39 & -16.3 & -1.0 &  \\ 
45 & J1001-06 & MCG-01-26-009* & 10 01 33.6 & -06 31 30 & 748 & 8.2 & 38.4 & 7.78 & -14.3 & -0.5 & ii \\ 
46 & J1039+01 & UGC05797* & 10 39 25.2 & +01 43 07 & 711 & 7.2 & 47.8 & 7.95 & -15.3 & -2.1 & iii \\ 
47 & J1103-34 & ESO377-G003 & 11 03 55.2 & -34 21 30 & 998 & 11.1 & 52.6 & 8.22 & -15.5 & -1.1 &  \\ 
48 & J1107-17 & 2MASXJ11070378-1736223* & 11 07 03.8 & -17 36 22 & 993 & 11.9 & 110 & 8.26 & -15.7 & -2.0 & ii \\ 
49 & J1111-24 & ESO502-G023 & 11 12 13.8 & -24 13 55 & 1455 & 19.9 & 102.5 & 8.40 & -16.5 & 0.0 &  \\ 
50 & J1118-17 & HIPASS J1118-17* & 11 18 03.1 & -17 38 31 & 1068 & 13.5 & 56.6 & 8.56 & -13.5 & -2.2 & iv \\ 
51 & J1137-39 & ESO320-G014 & 11 37 53.2 & -39 13 13 & 654 & 6.0** & 40.2 & 7.74 & -13.7 & -2.4 & \\ 
52 & J1142-19 & ESO571-G018 & 11 42 50.9 & -19 04 03 & 1415 & 20.0 & 111.5 & 8.31 & -16.7 & -1.7 & \\ 
53 & J1150-12 & MCG-02-30-033* & 11 50 36.4 & -12 28 04 & 1278 & 19.0 & 97.1 & 8.36 & -16.5 & -1.0 & ii \\ 
54 & J1223-13 & UGCA278* & 12 23 10.3 & -13 56 45 & 1163 & 16.0 & 83.3 & 8.27 & -16.1 & -1.8 & iii \\ 
55 & J1244-35 & ESO381-G018$\dagger$ & 12 44 42.4 & -35 58 00 & 625 & 5.3** & 39.8 & 7.84 & -13.4 & -1.3 & \\  
56 & J1259-19 & SGC1257.3-1909* & 12 59 56.3 & -19 24 47 & 826 & 8.5 & 47.2 & 8.14 & -13.5 & -1.8 & iii \\ 
57 & J1305-40 & [KK98]182 & 13 05 02.1 & -40 04 58 & 620 & 5.8** & 38 & 8.08 & -12.7 & 0.0 & \\ 
58 & J1308-16 & MCG-03-34-002* & 13 07 56.6 & -16 41 21 & 959 & 11.4 & 56.6 & 8.07 & -16.0 & -1.2 & I \\ 
59 & J1309-27 & AM1306-265* & 13 09 36.6 & -27 08 27 & 684 & 6.2 & 58.6 & 8.19 & -14.9 & -0.6 & iii \\ 
60 & J1322-28 & 6dF J1322443-285711* & 13 22 44.3 & -28 57 11 & 990 & 11.7 & 102.5 & 8.25 & -15.4 & -3.6 & iii \\ 
61 & J1337-28 & ESO444-G084* & 13 37 20.0 & -28 02 42 & 587 & 5.1** & 58.9 & 8.62 & -14.2 & 0.0 & iii \\
62 & J1349-12 & HIPASS J1349-12* & 13 49 10.0 & -12 45 35 & 1395 & 21.3 & 90.2 & 8.50 & -16.4 & -0.3 & iii \\ 
63 & J1350-35 & ESO383-G092 & 13 50 42.0 & -35 54 55 & 1411 & 17.2 & 46 & 8.38 & -16.5 & -0.4 &  \\ 
64 & J1355-23 & ESO510-G015* & 13 55 03.3 & -23 12 54 & 1372 & 18.5 & 65.2 & 8.43 & -15.5 & -0.9 & iii+iv \\ 
65 & J1403-27 & ESO510-IG052* & 14 03 34.6 & -27 16 47 & 1326 & 17.5 & 119.8 & 8.72 & -16.6 & -1.8 & I \\ 
66 & J1443-33 & ESO386-G013 & 14 43 04.3 & -33 29 26 & 1378 & 17.8 & 93.1 & 8.33 & -15.5 & -1.5 &  \\ 
67 & J1609-04 & MCG-01-41-006* & 16 09 36.8 & -04 37 13 & 829 & 14.8 & 71.9 & 8.30 & -16.1 & -2.9 & ii \\ 
68 & J2003-31 & ESO461-G036 & 20 03 51.0 & -31 41 53 & 428 & 7.8** & 71.6 & 8.02 & -14.1 & - & \\ 
69 & J2039-63 & 2MASXJ20385728-6346157* & 20 38 57.2 & -63 46 16 & 1656 & 22.8 & 49.2 & 8.31 & -16.5 & -1.4 & I \\ 
70 & J2056-16 & HIPASS J2056-16*$\dagger$ & 20 56 51.0 & -16 30 37 & 1453 & 22.8 & 66.7 & 8.38 & -16.2 & -0.1 & iii \\ 
71 & J2142-06 & 6dFJ2142269-061954* & 21 42 26.9 & -06 19 58 & 1249 & 19.9 & 62.7 & 8.26 & -16.5 & -2.0 & iii \\ 
72 & J2207-43 & ESO288-IG042* & 22 07 50.9 & -43 16 43 & 1396 & 20.0 & 58.4 & 8.51 & -16.4 & -2.5 & iii \\ 
73 & J2234-04 & MCG-01-57-015* & 22 34 54.7 & -04 42 04 & 889 & 14.1 & 92.9 & 8.50 & -16.2 & -0.2 & iii \\ 
74 & J2239-04 & UGCA433* & 22 39 09.0 & -04 45 37 & 831 & 12.0** & 57.5 & 8.37 & -16.0 & -1.2 & iii \\ 
75 & J2242-06 & 6dFJ2242235-065010*$\dagger$ & 22 42 23.5 & -06 50 10 & 899 & 14.1 & 62.6 & 7.95 & -15.6 & -0.7 & ii \\ 
76 & J2254-26 & MCG-05-54-004* & 22 54 45.2 & -26 53 25 & 819 & 12.1 & 126.6 & 8.46 & -16.1 & -2.1 & I \\ 
77 & J2255-34 & ESO406-G022* & 22 55 52.6 & -34 33 18 & 1286 & 18.2 & 66.8 & 8.16 & -16.6 & -0.6 & ii \\ 
78 & J2259-13 & APMUKS(BJ)B225708.22-133928.9 & 22 59 46.2 & -13 23 22 & 1219 & 17.9 & 54.9 & 7.99 & -15.9 & -4.0 &  \\ 
79 & J2311-42 & ESO291-G003 & 23 11 10.9 & -42 50 51 & 1381 & 19.1 & 94 & 8.19 & -16.5 & -1.3 &  \\ 
80 & J2334-45b & ESO291-G031* & 23 34 20.8 & -45 59 50 & 1487 & 20.5 & 89.7 & 8.42 & -16.0 & -0.6 & iii \\ 
81 & J2349-22 & APMUKS(BJ)B234716.77-224937.5 & 23 49 51.8 & -22 32 56 & 1020 & 14.6 & 69.7 & 7.99 & -14.7 & -3.0 &  \\ 
82 & J2349-37 & ESO348-G009* & 23 49 23.5 & -37 46 19 & 647 & 7.7** & 84.9 & 8.47 & -15.1 & -0.5 & iii \\ 
83& J2352-52 & ESO149-G003* & 23 52 02.8 & -52 34 40 & 574 & 6.2** & 47.9 & 8.13 & -14.9 & -6.8 & ii \\ 
\enddata
\tablecomments{
(1) Objects marked ($\dagger$) have no GALEX image\\
(2) Objects marked (*) were observed in the SINGG program\\
(3) Distances marked (**) are directly measured values from the literature, as reported in NED; all others are calculated from redshifts and the flow model.\\
(4) The `Type' category classifies the morphology of the H$\alpha$ regions and is described in $\S5$. It is only evaluated for objects with SINGG images, where information on the H$\alpha$ regions is available.}
\end{deluxetable}

\newpage

\begin{deluxetable}{lrrccccc}
\tabletypesize{\scriptsize}
\tablecolumns{8} 
\tablewidth{0pc} 
\tablecaption{Values used for projected cluster exclusion radii and velocity dispersions\label{table_2}} 
\tablehead{ 
\colhead{Cluster} & \colhead{RA} & \colhead{\hskip -0.2in (J2000) Dec}  & \colhead{V$_h$} & \colhead{Excluded radius} & \colhead{$\sigma$} & \colhead{Excluded} \\[-1ex]
\colhead{} & \colhead{(deg)} & \colhead{(deg)} & \colhead{(km s$^{-1}$)} & \colhead{(\degree)} & \colhead{(km s$^{-1}$)} & \colhead{}  }
\startdata 

Virgo & 186.75 & 12.72 & 1059 & 25 & 757 & 14 \\ 
Fornax & 45.63 & -35.45 & 1583 & 15 & 429  & 11 \\ 
Eridanus & 54.50 & -22.32 & 1657 & 10 & 179 & 1 \\ 
Antlia & 157.50 & -35.32 & 2797 & 7 & 469 & 2 \\ 
Cen30 & 192.00 & -51.80 & 3397 & 12 & 933 & 1 \\ 
Hydra & 150.61 & -27.53 & 3777 & 7 & 608 & 1 \\
\enddata
\tablecomments{The Exclusion Radius figures are based where available on best estimates of the zero-infall radius.  Where the zero-infall radius is at best known only poorly, a conservative estimate has been adopted. $\sigma$ is the cluster velocity dispersion. `Excluded' indicates the number of candidates that would otherwise be in the sample that are within the exlusion radii. All cluster center values are approximate.}
\end{deluxetable}

\newpage

\begin{deluxetable}{clllcrrcc} 
\tabletypesize{\scriptsize}
\tablecolumns{9} 
\tablewidth{0pc} 
\tablecaption{Comparison of tidal indices\label{table_3}} 
\tablehead{ 
\colhead{SIGRID} & \colhead{Object} & \colhead{RA} & \colhead{\hskip -0.3in (J2000) \quad Dec}  & \colhead{V$_h$} & \colhead{$\Theta$} & \colhead{$\Delta$} & \colhead{T$_{last}$} & \colhead{Disturbers} \\
\colhead{ID} & \colhead{}   & \colhead{(h m s)}   & \colhead{\hskip 0.15in(d m s)} & 
\colhead{(km s$^{-1}$)}    & \colhead{tidal}   & \colhead{}    & \colhead{(Gyr)} & \colhead{($<$ 15Gyr)}}
\startdata 
5 & IC1574 & 00 43 03.8 & -22 15 01 & 363 & -0.1 & --1.0 & 5.7 & 3 \\ 
36 & ESO121-G020 & 06 15 54.5 & -57 43 35 & 583 & -1.6 & -1.9 & 9.6 & 1 \\ 
   & ESO489-G?056 & 06 26 17.0 & -26 15 56 & 492 & -2.1 & -0.7 & 3.6 & 1 \\ 
   & ESO308-G022 & 06 39 32.7 & -40 43 15 & 821 & -2.6 & -1.8 & 20 & 0 \\
51 & ESO320-G014 & 11 37 53.4 & -39 13 14 & 654 & -1.2 & -0.9 & 10.6 & 3 \\ 
   & ESO379-G007 & 11 54 43.0 & -33 33 29 & 640 & -1.3 & -1.0 & 4.9 & 3 \\ 
   & ESO321-G014 & 12 13 49.6 & -38 13 53 & 615 & -0.3 & -1.6 & 5.0 & 5 \\ 
55 & ESO381-G018 & 12 44 42.7 & -35 58 00 & 625 & -0.6 & -1.3 & 6.0 & 6 \\ 
57 & [KK98]182 & 13 05 02.9 & -40 04 58 & 620 & 1.2 & 0.0 & 6.5 & 5 \\ 
59 & AM1306-265 & 13 09 36.6 & -27 08 26 & 684 & -0.6 & -0.6 & 7.9 & 5 \\ 
   & [KK98]195 & 13 21 08.2 & -31 31 47 & 567 & -0.2 & 0.2 & 5.6 & 15 \\ 
   & UGCA365 & 13 36 30.8 & -29 14 11 & 570 & 2.1 & 0.4 & 2.9 & 11 \\ 
66 & ESO444-G084 & 13 37 20.2 & -28 02 46 & 587 & 1.7 & 0.0 & 2.6 & 11 \\ 
   & HIPASSJ1337-39 & 13 37 25.1 & -39 53 52 & 492 & -0.3 & 0.4 & 4.4 & 11 \\ 
   & ESO272-G025 & 14 43 25.5 & -44 42 19 & 631 & -1.5 & -1.8 & 7.6 & 2 \\ 
83 & ESO149-G003 & 23 52 02.8 & -52 34 39 & 574 & -1.7 & -6.7 & - & 0 \\ \\
\enddata
\tablecomments{Notes: V$_h$ is the target galaxy's heliocentric recession velocity and T$_{last}$ is the most recent possible interaction time between the object and its possible disturbers, assuming a separation velocity of 100 km~s$^{-1}$ for each object pair.  The disturber index $\Delta$ is the largest value of the index for the set of disturbers identified from NED.  $\Delta$ is thus arbitrary, but, like $\Theta$, indicates isolation when less than zero.}

\end{deluxetable} 

\end{document}